# THE ROTATION AND OTHER PROPERTIES OF COMET 49P/AREND-RIGAUX, 1984 – 2012

Nora Eisner[1,2], Matthew M. Knight[2], and David G. Schleicher[3]

[1] Department of Physics and Astronomy, University of Sheffield, Sheffield, S3 7RH; nora.eisner@hotmail.com
[2] Department of Astronomy, University of Maryland, College Park, MD 20742, USA
[3] Lowell Observatory, 1400 W. Mars Hill Rd, Flagstaff, AZ 86001, USA

## ABSTRACT

We analyzed images of comet 49P/Arend-Rigaux on 33 nights between 2012 January and May and obtained *R*-band lightcurves of the nucleus. Through usual phasing of the data we found a double-peaked lightcurve having a synodic rotation period of 13.450 ± 0.005 hr. Similarly, phase dispersion minimization and the Lomb-Scargle method both revealed rotation periods of 13.452 hr. Throughout the 2011/12 apparition, the rotation period was found to increase by a small amount, consistent with a retrograde rotation of the nucleus. We also reanalyzed the publicly available data from the 1984/85 apparition by applying the same techniques, finding a rotation period of 13.45 ± 0.01 hr. Based on these findings we show that the change in rotation period is less than 14 seconds per apparition. Furthermore, the amplitudes of the light curves from the two apparitions are comparable, to within reasonable errors, even though the viewing geometries differ, implying that we are seeing the comet at a similar sub-Earth latitude. We detected the presence of a short term jet-like feature in 2012 March which appears to have been created by a short duration burst of activity on March 15. Production rates obtained in 2004/05, along with reanalysis of previous results from 1984/85 imply a strong seasonal effect and a very steep fall-off after perihelion. This, in turn, implies that a single source region dominates activity, rather than leakage from the entire nucleus.

## 1. INTRODUCTION

Comets spend most of their lifetimes in the cold outer Solar System and are therefore believed to be largely unchanged since the era of planetary formation (e.g., Mumma et al. 1993, Dones et al. 2004). This makes them ideal tools for studying the early conditions of the Solar System as well as properties of the protoplanetary disk. Furthermore, their physical properties must be explained by any unified theory of the evolution of the Solar System, and thus they are valuable for testing such theories. For a complete understanding of comets, one requires knowledge of the orbital path, rotational period and activity as these properties are all closely linked. In order to correctly interpret observations of the coma and determine nuclear activity across the surface, for example, it is essential to know the rotation period of the cometary nucleus (e.g., Samarasinha et al. 2004). Furthermore, the rotation period can help infer information about the comet's internal structure, as has been confirmed by occasional spacecraft visits (e.g., Barucci et al. 2011).

The rotation period of a comet can be determined by analyzing its lightcurve. The necessary photometric measurements, however, are often difficult to obtain due to contributions from the cometary coma, which scatters light and thus overwhelms the light reflected off the nucleus. Due to this, ground based observations of cometary nuclei are limited to comets at large heliocentric distances or to comets which are relatively 'anemic' in their production of the coma. The latter means that observations can take place when the comet is close to the Earth, and small photometric apertures can be used to reduce the coma contribution.

The first nucleus lightcurves of an anemic comet were obtained in 1984 of comet 28P/Neujmin 1 (A'Hearn et al. 1984a; Campins et al. 1987). Observations were carried out in both the optical and the thermal IR; however, without complete lightcurves in either of these wavebands they were unable to unambiguously state the cause of the variations in brightness. Soon thereafter, the anemic comet 49P/Arend-Rigaux was observed. As was done for comet 28P/Neujmin 1, Millis et al. (1985;1988) observed comet 49P/Arend-Rigaux in the optical as well as in the thermal IR. In the case of 49P/Arend-Rigaux, the observations in the thermal IR were sufficient to allow them to conclude that the variations in optical brightness were due to the shape of the nucleus as opposed to changes in albedo across the surface. The thermal IR data also confirmed the shape of the nucleus to be that of a near triaxial ellipsoid with dimension of 13 × 8 × 8 km, resulting in a double-peaked lightcurve. The thermal IR measurements of both of these objects revealed that comet nuclei have extremely low albedos before this was 'discovered' by the 1986 spacecraft visit to 1P/Halley (Keller et al. 1986).

Comet 49P/Arend-Rigaux was observed and analyzed independently by three groups during its 1984/85 apparition: Jewitt & Meech (1985), Millis et al. (1988) and Wisniewski et al. (1986). All three groups concluded different values for the nucleus rotation period and although all values were simple ratios of one another, no individual set of data had sufficient observations to allow for high levels of precision or the complete removal of aliases. By combining the optical data from these three independent groups, we increased the total number of nights of observations and thus were able to obtain a more precise value of the rotation period during the 1984/85 apparition. The calculated value was compared to the rotation period obtained for the most recent 2011/12 apparition, which reached perihelion on 2011 October 19.1. During this apparition we observed comet 49P/Arend-Rigaux from 2012 January until May, obtaining images in the broadband *R*-filter to measure the nucleus lightcurve. Further observational data were collected during the 2004 apparition, however, the data were of poor quality and proved to be unusable for constraining the rotation period.

The direct comparison between the 1985 and 2012 data enabled us to determine whether there was a significant change in rotation period between the two apparitions. Computational models suggest that changes in rotation periods should be common, nonetheless, this has only been conclusively demonstrated in a small number of comets, e.g., Comet Levy (Feldman et al. 1992; Schleicher et al. 1991), 10P/Tempel 2 (Mueller & Ferrin 1996; Knight et al. 2011, 2012; Schleicher et al. 2013), 2P/Encke (Fernández et al. 2005), 9P/Tempel 1 (Chesley et al. 2013), 103P/Hartley 2 (e.g., Knight & Schleicher 2011; Samarasinha et al. 2011; Knight et al. 2015), and 41P/Tuttle-Giacobini-Kresak (Bodewits et al. 2017; Knight et al. 2017). Furthermore, recent observations from the *Rosetta* spacecraft showed that the rotation period of 67P/Churyumov-Gerasimenko changed throughout the orbit, with an increase in rotation period of 0.2% as it approached perihelion followed by a rapid decrease of 1% as it moved further away (e.g., Keller et al. 2015; Jorda et al. 2016). The lack of more observational evidence for changes in rotation period is assumed to be largely due to the lack of high quality data for multiple apparitions of the same comet.

The most common cause for changes in the rotation period of a comet is believed to be asymmetric outgassing resulting in torquing (e.g., Samarasinha et al. 2004). This suggests that comets with a smaller nucleus are more prone to changes in rotation period and thus that the large nucleus of comet 49P/Arend-Rigaux is unlikely to undergo rapid changes. Furthermore, the 2012 observations showed very low levels of outgassing and, despite efforts to enhance the images (e.g., Schleicher & Farnham 2004), we were unable to detect any morphological evidence of dust jets which could affect the rotation period. Nonetheless, the stacking of nightly images revealed the presence of a short term jet-like feature in 2012 March, which will be discussed later. The comet was too faint for our standard narrowband imaging and therefore it is unknown if any gas jets exist.

Likewise, the comet was too faint for our standard photoelectric photometer observations in 2011/2012; however, we were able to obtain data during the 2004/05 apparition using a photoelectric photometer with narrowband comet filters. Similar observations were carried out during the 1984/85 apparition (Millis et al. 1988), thus allowing us to compute and intercompare production rates and abundance ratios of a number of gas species from these two apparitions.

The layout of the paper is as follows. A summary of the observations and reductions of the 2012 imaging data is found in *Section 2*, followed by an in depth analysis of the light curve in *Section 3*. *Section 4* explains and analyzes a number of properties of the coma and the final section provides an overall summary and discussion of all the results.

## 2. OBSERVATIONS AND REDUCTIONS OF IMAGING IN 2012

### 2.1. Observing Overview

Useful images of comet 49P/Arend-Rigaux were obtained on a total of 33 nights between 2012 January and May with sampling at monthly intervals (*Table 1*). Observations were obtained at the Lowell Observatory Hall 1.1 m telescope with the e2v CCD231-84. On-chip 2 × 2 binning produced images with a pixel scale of 0.740 arcseconds pixel$^{-1}$. On-chip 3 × 3 binning was used for observations in May, producing images with a pixel scale of 1.11 arcseconds pixel$^{-1}$. The images obtained with the 1.1 m telescope were guided at the comet's rate of motion, with the exception of the data collected in May, which were trailed at half the comet's rate, resulting in equal trailing of the stars and the comet. Additional observations were obtained with the 0.8 m telescope, also at Lowell Observatory, with the e2v CCD42-40. On-chip 2 × 2 binning produced images with a pixel scale of 0.456 arcseconds pixel$^{-1}$. The images obtained with the 0.8 m telescope were guided at the sidereal rate, with the exception of the first three nights in January, which were tracked at the comet's ephemeris rate. Broadband *R*-filters were used for all observations except those carried out in May, which used the *VR*-filter (about twice as wide as a standard *R*-filter) in order to improve the signal-to-noise. Exposure times prior to 2012 March 21 were typically 120 seconds; exposure times thereafter were always 300 seconds. The large variety of techniques used for these observations are due to individual observational runs being carried out by different observers.

### 2.2. Absolute Calibrations

The data were reduced using standard techniques in IDL to remove bias and apply flat fields (e.g., Knight & Schleicher 2015). Landolt standard stars (Landolt 2009) were observed to determine the instrumental magnitude and extinction coefficients on 2012 January 25 and 26 for the 0.8 m and 1.1 m telescope respectively (although *Table 1* shows these nights as "cirrus", only parts of these nights had cirrus and it was photometric when the standard stars were observed). Standard stars were not observed on other nights so typical zero-point and extinction coefficients were used. The application of absolute calibrations on non-photometric nights provided first order corrections for airmass, which allowed us to determine the additional offsets necessary due to clouds. We confirmed that these produced reasonable calibrations by spot-checking selected dates against UCAC5 *R*-filter catalog values (Zacharias et al. 2017) for a number of field stars, finding typical photometric accuracies of <0.1 mag. As will be discussed in *Section 2.6*, additional small nightly offsets were applied in order to align the light curves.

We are not aware of any calibration fields for the *VR*-filter (used in May) so have treated these images like *R*-band images. This likely resulted in a small (<0.05) tilt to fainter magnitudes at high airmass due to a different extinction correction and larger (0.1-0.3 mag) absolute calibration offsets to brighter magnitudes, as reflected in the $\Delta m_2$ values given in *Table 1*. As a result, the May data were used only for period determination since this was unaffected by the calibration issues.

### 2.3. Comet Measurements

The flux of the comet was determined by centroiding on the nucleus and integrating inside circular apertures. Similarly, the median sky flux was calculated in an annulus centered on the nucleus, with a radius large enough to avoid coma contamination. Apertures from 3 to 30 pixels in radius were

**Table 1**
Summary of Comet 49P/Arend-Rigaux Observations and Geometric Parameters during our **2012** Observations[a]

| UT Date | UT Range | ΔT (days)[b] | Tel. Diam. | $r_H$ (AU) | Δ (AU) | PA (°)[c] | α (°)[d] | Δt (hr)[e] | $\Delta m_1$[f] | $\Delta m_2$[g] | $\sigma_m$[h] | Ap. Rad. (")[i] | Conditions |
|---|---|---|---|---|---|---|---|---|---|---|---|---|---|
| Jan 25 | 06:43-08:09 | 98.22  | 0.8 m | 1.774 | 1.034 | 102.0 | 27.7 | 0.214 | -2.425 | -0.101 | 0.0054 | 3.6 | Cirrus |
| Jan 26 | 06:51-13:33 | 99.34  | 0.8 m | 1.781 | 1.032 | 101.3 | 27.3 | 0.143 | -2.414 | 0.000  | 0.0048 | 3.2 | Cirrus |
| Jan 26 | 06:33-10:50 | 99.33  | 1.1 m | 1.780 | 1.032 | 101.3 | 27.3 | 0.143 | -2.412 | -0.005 | 0.0034 | 1.6 | Cirrus |
| Jan 27 | 07:17-13:29 | 100.34 | 0.8 m | 1.787 | 1.031 | 100.6 | 26.9 | 0.143 | -2.403 | 0.045  | 0.0051 | 3.6 | Cirrus |
| Jan 27 | 06:16-12:15 | 100.33 | 1.1 m | 1.787 | 1.031 | 100.7 | 26.9 | 0.143 | -2.403 | 0.054  | 0.0039 | 2.0 | Cirrus |
| Jan 28 | 06:53-14:04 | 101.39 | 0.8 m | 1.793 | 1.029 | 99.9  | 26.5 | 0.143 | -2.390 | -0.102 | 0.0048 | 3.6 | Cirrus |
| Jan 29 | 06:58-13:38 | 102.34 | 0.8 m | 1.799 | 1.028 | 99.2  | 26.1 | 0.142 | -2.379 | 0.047  | 0.0049 | 3.6 | Cirrus |
| Jan 30 | 07:00-13:39 | 103.35 | 0.8 m | 1.805 | 1.027 | 98.5  | 25.6 | 0.142 | -2.364 | 0.078  | 0.0045 | 3.6 | Cirrus |
| Jan 30 | 05:58-09:04 | 103.32 | 1.1 m | 1.804 | 1.028 | 98.5  | 25.7 | 0.142 | -2.369 | 0.035  | 0.0036 | 1.6 | Cirrus |
| Feb 17 | 05:01-13:47 | 121.32 | 0.8 m | 1.919 | 1.038 | 78.0  | 18.2 | 0.144 | -2.224 | -0.097 | 0.0043 | 3.6 | Intermittent clouds |
| Feb 18 | 04:44-13:16 | 122.30 | 0.8 m | 1.926 | 1.040 | 76.3  | 17.8 | 0.144 | -2.220 | 0.018  | 0.0042 | 4.1 | Cirrus |
| Feb 19 | 04:38-07:44 | 123.17 | 0.8 m | 1.931 | 1.042 | 74.9  | 17.5 | 0.144 | -2.218 | -0.115 | 0.0058 | 3.6 | Intermittent clouds |
| Feb 21 | 04:34-13:13 | 125.29 | 0.8 m | 1.945 | 1.048 | 71.0  | 16.7 | 0.145 | -2.214 | -0.039 | 0.0048 | 3.6 | Cirrus |
| Feb 21 | 04:21-05:44 | 125.27 | 1.1 m | 1.944 | 1.047 | 71.3  | 16.8 | 0.145 | -2.215 | -0.040 | 0.0025 | 2.0 | Cirrus |
| Feb 22 | 03:56-12:05 | 126.24 | 1.1 m | 1.950 | 1.050 | 69.2  | 16.5 | 0.146 | -2.216 | -0.038 | 0.0026 | 2.0 | Cirrus |
| Feb 23 | 04:18-13:10 | 127.29 | 0.8 m | 1.958 | 1.054 | 67.1  | 16.1 | 0.146 | -2.217 | -0.027 | 0.0040 | 3.6 | Cirrus |
| Feb 24 | 04:12-13:10 | 128.28 | 0.8 m | 1.965 | 1.058 | 65.0  | 15.8 | 0.147 | -2.221 | -0.059 | 0.0044 | 4.1 | Photometric |
| Feb 25 | 04:07-13:09 | 129.28 | 0.8 m | 1.971 | 1.061 | 62.9  | 15.5 | 0.147 | -2.222 | 0.064  | 0.0042 | 4.1 | Cirrus |
| Feb 26 | 04:02-13:05 | 130.28 | 0.8 m | 1.978 | 1.065 | 60.7  | 15.2 | 0.148 | -2.226 | 0.018  | 0.0039 | 4.1 | Photometric |
| Mar 16 | 02:25-11:55 | 149.25 | 1.1 m | 2.105 | 1.180 | 12.0  | 13.4 | 0.164 | -2.512 | 0.293  | 0.0024 | 2.0 | Cirrus |
| Mar 21 | 02:43-12:42 | 154.23 | 0.8 m | 2.139 | 1.224 | 0.2   | 14.0 | 0.170 | -2.650 | 0.230  | 0.0028 | 4.6 | Photometric |
| Mar 22 | 02:44-12:34 | 155.23 | 0.8 m | 2.146 | 1.233 | 358.0 | 14.1 | 0.171 | -2.677 | 0.203  | 0.0027 | 4.6 | Photometric |
| Mar 25 | 02:47-12:33 | 158.18 | 0.8 m | 2.166 | 1.262 | 352.0 | 14.6 | 0.175 | -2.768 | 0.090  | 0.0033 | 3.6 | Cirrus |
| Mar 27 | 02:48-12:26 | 160.24 | 0.8 m | 2.180 | 1.283 | 348.2 | 15.0 | 0.178 | -2.833 | 0.033  | 0.0028 | 4.6 | Intermittent clouds |
| Mar 28 | 02:49-12:28 | 161.24 | 0.8 m | 2.187 | 1.293 | 346.6 | 15.2 | 0.179 | -2.865 | 0.020  | 0.0027 | 4.6 | Cirrus |
| Apr 17 | 02:46-10:28 | 181.17 | 1.1 m | 2.324 | 1.541 | 321.6 | 19.1 | 0.214 | -3.534 | -0.094 | 0.0021 | 2.9 | Cirrus |
| Apr 18 | 03:05-10:46 | 182.20 | 0.8 m | 2.330 | 1.556 | 320.7 | 19.3 | 0.216 | -3.569 | 0.038  | 0.0040 | 3.6 | Cirrus |
| Apr 18 | 03:59-10:18 | 182.21 | 1.1 m | 2.330 | 1.556 | 320.7 | 19.3 | 0.216 | -3.569 | 0.030  | 0.0028 | 2.3 | Cirrus |
| Apr 19 | 03:06-10:44 | 183.20 | 0.8 m | 2.337 | 1.569 | 319.8 | 19.5 | 0.217 | -3.601 | 0.032  | 0.0037 | 4.1 | Cirrus |
| Apr 19 | 02:47-10:12 | 183.18 | 1.1 m | 2.337 | 1.569 | 319.8 | 19.5 | 0.217 | -3.601 | -0.010 | 0.0037 | 2.3 | Intermittent clouds |
| Apr 20 | 03:07-10:40 | 184.20 | 0.8 m | 2.344 | 1.584 | 319.0 | 19.6 | 0.220 | -3.633 | -0.060 | 0.0044 | 3.2 | Intermittent clouds |
| Apr 21 | 03:08-10:35 | 185.21 | 0.8 m | 2.351 | 1.598 | 318.2 | 19.8 | 0.222 | -3.666 | 0.055  | 0.0039 | 3.6 | Intermittent clouds |
| Apr 22 | 03:09-10:30 | 186.19 | 0.8 m | 2.358 | 1.613 | 317.5 | 19.9 | 0.224 | -3.697 | 0.058  | 0.0040 | 3.6 | Intermittent clouds |
| Apr 23 | 03:14-10:27 | 187.19 | 0.8 m | 2.365 | 1.628 | 316.7 | 20.1 | 0.226 | -3.731 | 0.043  | 0.0041 | 3.6 | Intermittent clouds |
| Apr 24 | 03:15-10:23 | 188.23 | 0.8 m | 2.371 | 1.643 | 316.0 | 20.2 | 0.228 | -3.761 | -0.090 | 0.0039 | 3.6 | Cirrus |
| Apr 25 | 03:16-10:17 | 189.19 | 1.1 m | 2.378 | 1.658 | 315.3 | 20.4 | 0.230 | -3.795 | 0.065  | 0.0036 | 4.1 | Cirrus |
| May 12 | 03:32-07:03 | 206.18 | 1.1 m | 2.493 | 1.929 | 305.9 | 22.0 | 0.267 | -4.290 | 0.195  | 0.0016 | 2.6 | Photometric |
| May 13 | 03:10-08:44 | 207.13 | 1.1 m | 2.500 | 1.947 | 305.4 | 22.1 | 0.270 | -4.321 | 0.282  | 0.0017 | 2.6 | Intermittent clouds |
| May 14 | 03:18-08:33 | 208.15 | 1.1 m | 2.507 | 1.964 | 305.0 | 22.1 | 0.272 | -4.345 | 0.237  | 0.0018 | 2.6 | Photometric |

**Notes.**
[a] All parameters were taken at the midpoint of each night's observations, and all images were obtained at Lowell Observatory.
[b] Time since perihelion (2011 October 19.1).
[c] Position angle of the Sun.
[d] Solar phase angle.
[e] Light travel time.
[f] Magnitude necessary to correct for changes in the geometry ($\Delta m_1 = -5 \log(r_h \Delta) - \alpha \beta$).
[g] Offset (in magnitudes) necessary to make data on all nights peak at the same magnitude, after correcting for geometry.
[h] Average uncertainty calculated for the night.
[i] Radius of the aperture used to extract the lightcurves (see *Section 2.3*).

used for the comet, allowing us to monitor for passing stars, which showed up in the larger apertures first. The aperture with the most coherent lightcurve was independently determined for each night (given in column 13 of *Table 1*), on the basis that it had to be large enough to include as much light from the nucleus as possible but small enough to avoid contamination from passing stars. This depended on a variety of factors, including trailing and seeing, as well as how crowded the field was, but was generally ∼ 3 arcseconds, i.e. around twice the FWHM.

### 2.4. Comparison Star Correction

Following the methodology of earlier papers (e.g., Knight et al. 2011), we conducted photometry on seven field stars per night, allowing us to correct for transparency variations or changes in the sensitivity of the equipment. The magnitude of each field star was tracked throughout the night using the same range of apertures as used for the comet. As the field stars are expected to maintain a constant brightness in good weather conditions, any deviation from this least obscured brightness suggests a change in observing conditions. The fourth brightest measurement for each star was used as the least obscured brightness because it is statistically unlikely for more than four frames to be affected by random fluctuations such as cosmic rays, bad pixels etc. which might bias occasional pixels too high. The corrections necessary to bring fainter measurements into agreement with this value were determined for each frame and the median offset of the seven field stars calculated to give a correction value for each image. These magnitude corrections were then applied to the comet's magnitude in that image, yielding a corrected lightcurve, as shown in *Figure 1*. This method is based on the assumption that the conditions were photometric at least once during the night. If this was not the case, then the night in question will be systematically fainter and will have a correspondingly smaller (more negative) $\Delta m_2$, as tabulated in column 11 of *Table 1*. As discussed in *Section 2.6*, all $\Delta m_2$ values were within 0.12 mag of 0.0 except for nights with known reasons (outburst, different filter), dust was released near perihelion, when the dust activity was at its greatest (see *Section 4.2*). Even though the rotation period can be obtained without removal of the coma, as will be discussed, in order to accurately determine the amplitude (peak-to-trough variation) of the lightcurve and to compare results to those of other authors, the extent of the coma contribution was considered.

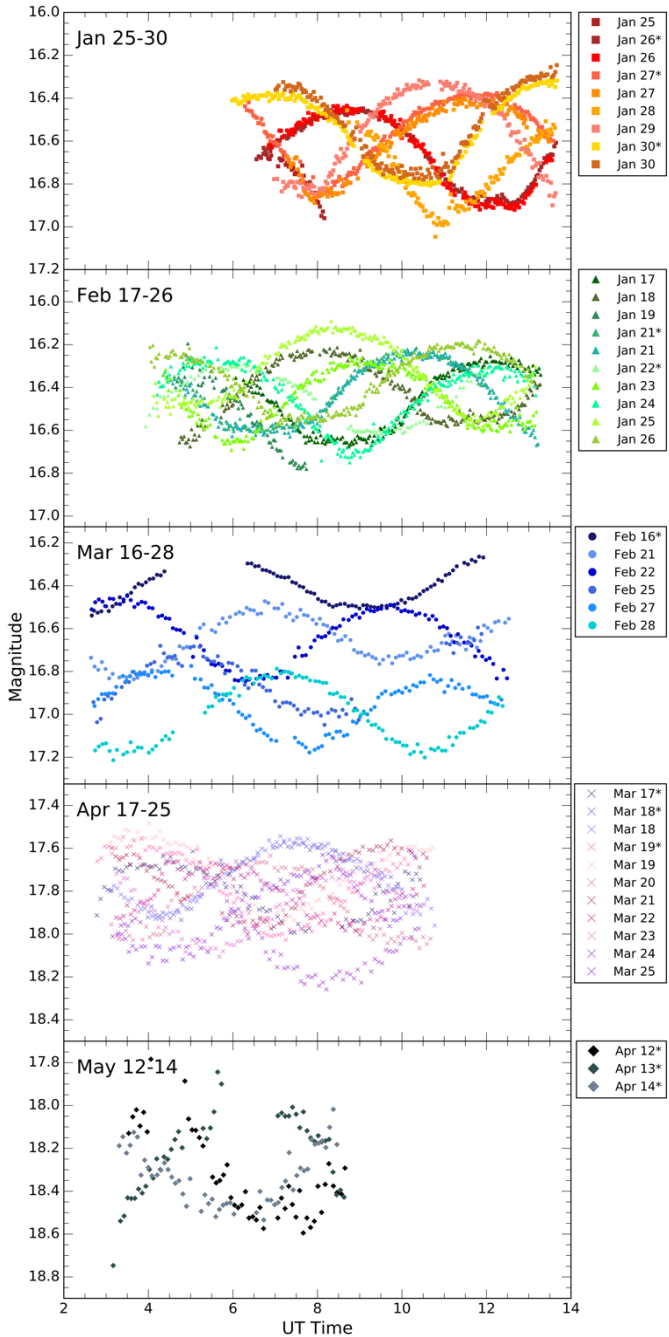

**Figure 1.** Our observed *R*-band magnitudes (*m*) plotted as a function of UT on each night. The magnitudes have had absolute calibration, extinction correction and field star corrections applied. The right hand panels describe the symbol for each night, where the dates with an asterix (*) show data collected with the 1.1 m telescope, whilst the remaining data were obtained with the 0.8 m telescope.

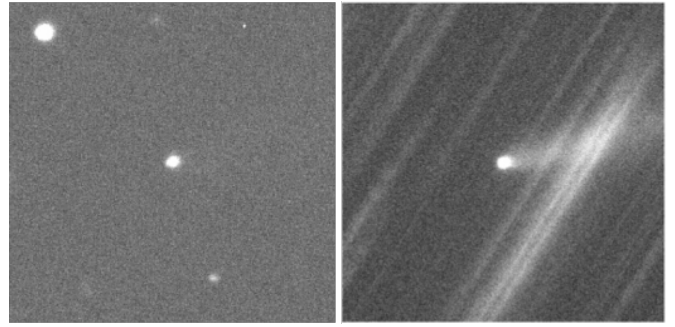

**Figure 2.** Single frame (left) and stacked image (right) from 2012 February 26. The tail is only clearly visible in the stacked image. Both images represent a physical size of 242,000 km across at the comet.

The coma contribution was calculated following the commonly used coma removal method of e.g., Millis et al. (1988) and Knight et al. (2011) for images tracked at the comet's rate. This method is based on the assumption that the dust grains move out from the nucleus isotropically at a constant velocity and thus that the coma flux per pixel decreases as $\rho^{-1}$, where $\rho$ is the radial distance from the nucleus. Conversely, the area of equally spaced annuli increases as $\rho$, and therefore the total coma flux in each annulus should be constant. Even though many factors can influence the validity of this assumption, such as contamination from field stars or cosmic rays, it has been shown that a linear fit to the total annular flux as a function of radial distance gives a good first order approximation for the coma contribution.

The total flux was calculated in 3 pixel wide annuli centered on the nucleus ranging from 3 to 30 pixels (i.e. 3-6 pixels, 6-9 pixels,…, 27-30 pixels). These values were used to create a radial profile (total flux in annulus as a function of $\rho$), for each image over the course of a night (solid grey lines in *Figure 3*). A straight line was fit to the total annular flux as a function of radial distance for radii chosen to begin beyond significant nucleus signal. This threshold value varied with pixel scale and seeing.

When determining the coma contribution, frames with obvious star contamination were omitted on the basis of having a considerably higher than average flux at large $\rho$ (dashed blue lines in *Figure 3*). Stars at large radial distances would have resulted in an underestimate of the coma whilst stars at small radial distances would have resulted in an overestimate of the coma. Even though the stars with large contamination were removed, fainter stars will still be present. Nonetheless, the median combination of a large number of images minimizes their effect and produces an excellent fit to the coma as shown in *Figure 3*.

The nucleus flux was estimated for all nights that were guided at the comet's rate, by subtracting the modeled flux of the coma from the integrated flux within the photometric aperture for each frame. The nucleus and coma fluxes were then compared for all frames of a night to estimate the coma contamination. This revealed that the coma contribution on confirming that the comparison star correction technique worked as expected.

### 2.5. Coma Contamination

Comet 49P/Arend-Rigaux was already known to be one of the least active periodic comets (e.g., Jewitt & Meech 1985), nonetheless, there was some evidence of coma contamination throughout this apparition. Nightly observations were stacked and median combined into a single image in order to enhance the coma (see *Figure 2*). This revealed a persistent tail oriented due west from January through February, even though the position angle of the Sun changed from 102° to 60° over the same time span (column 7 of *Table 1*). The most likely explanation for this is that the nights with good seeing steadily decreased from 20% to 13% between January and May, as would be expected as the comet moved further away from the Sun. This result confirms that our photometry is dominated by nucleus signal. Nonetheless, nights with worse seeing conditions yield a lower percentage of nucleus flux because the PSF was worse, resulting in more nucleus signal falling outside of our photometric aperture, whilst the coma signal was calculated across larger apertures making it relatively impervious to seeing.

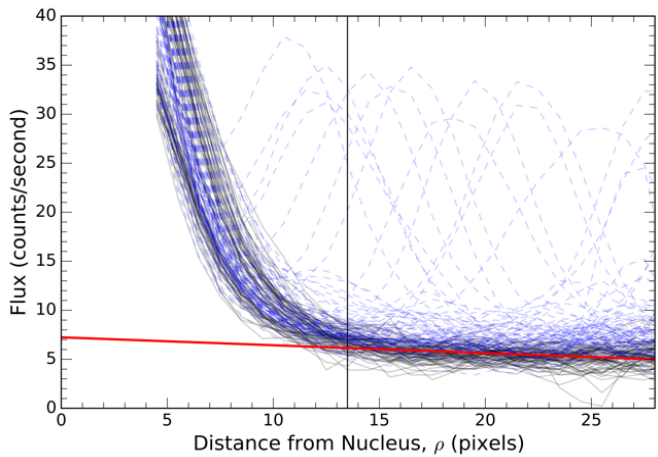

**Figure 3.** An example of a radial profile on 2012 January 30. Each curve represents the total annular flux in a 3 pixel wide annulus from images throughout the night. The solid grey curves are the frames which were used to calculate the nightly median coma profile (red solid line) whilst the dotted blue curves represent the frames which were disregarded due to star contamination. In this particular case, the coma was fit for annuli greater than 13.5 pixels in radius (the vertical line) and the nucleus lightcurve was extracted using an aperture radius of 8 pixels.

In order to determine whether the coma contribution varied over the course of a night we implemented the same methodology as described above for 30 minute time intervals. Although there were changes in the coma contribution during the course of the nights, there was not an obvious pattern in the variations. Due to the lack of evidence that the coma flux changed as a function of rotational phase, we decided not to remove the coma contribution. Furthermore, coma removal would introduce additional errors and would not likely improve our period determination.

### 2.6. Geometric Correction

As observations took place over several months, geometric effects had to be taken into account. The absolute magnitude (magnitude reduced to unit heliocentric and geocentric distances at zero solar phase angle), $M$, was found using standard asteroidal normalization (e.g., Jewitt 1991):

$$M = m_R - 5\log(r_H \Delta) - \alpha\beta \qquad [1]$$

where $m_R$ is the apparent magnitude (corrected for extinction and comparison stars as described in *Sections 2.2* and *2.4*), $r_h$ and $\Delta$ are the heliocentric and geocentric distances in AU respectively, $\alpha$ is the solar phase angle (Sun-comet-observer) and $\beta$ is the linear phase coefficient. The linear phase coefficient for comet nuclei has values 0.025 - 0.083 mag deg$^{-1}$ (Snodgrass et al. 2011). A value of 0.04 was adopted throughout. *Equation 1* corrects for the geometric variation in brightness and brings all the lightcurves to a similar scale. The geometric corrections ($\Delta m_1$ in column 10 of *Table 1*) were calculated at the midpoint of each night's observations, as the differences between the mid-points and the extremes on each night were usually comparable to the statistical uncertainty.

The geometric correction alone, however, was not sufficient to bring all the lightcurves from different nights to the same peak brightness. This could be due to a number of reasons, such as that our field star calibration was based on the assumption that each night was photometric at some point or due to changing levels of the comet's activity. Furthermore, the absolute calibrations assumed typical values for all but two nights. An additional adjustment was introduced in order to bring all the lightcurves to the same peak and is given as $\Delta m_2$ in column 11 of *Table 1*. This is simply a scaling factor to aid comparison of lightcurves and does not have physical significance. These corrections were always within 0.12 magnitudes of 0.0 except for March 16-22 when the brightness was enhanced following an outburst (discussed in *Section 4.1*) and in May 12-14 when the *VR*-filter was used and we could not perform absolute calibrations.

This additional correction factor also accounts for the difference in magnitudes due to nightly variations in the aperture radius. Whilst this variation could have also been minimized by using the same physical aperture size at the comet, we decided against doing this due to the generally worse seeing conditions at the 0.8 m telescope. The worse seeing would have required us to use a larger than optimum aperture on the 1.1 m telescope images, thus degrading these data. The extent of the effect of this is described in *Section 2.5*.

Whilst the additional correction factors and geometric correction helped to align the lightcurves in order to bring them to the same peak magnitude, they did not affect the time of the peaks and thus did not affect the rotational phasing. The data were, however, corrected for the time it took light to travel between the comet and us (column 9 in *Table 1*). Due to the change in Sun-comet-Earth geometry, this time differed by 0.13 hr over the course of our observations. The reduced magnitudes ($m_R^*$), as given in *Table 2*, have had absolute calibration, geometry, field star, and light travel time corrections, as well as $\Delta m_2$ offsets applied.

### 2.7. Further Corrections and Uncertainties

By close examination of the lightcurves, in conjunction with iterating through the nightly images, we identified frames which were contaminated by field stars, cosmic rays or tracking problems. These were discarded from the data set. A plot of the nightly field star correction values for each frame also helped to identify images with significant extinction due to clouds. Frames where the correction was larger than 0.5 magnitudes were individually reassessed in order to determine their utility, with the result of only a small number of frames being discarded.

Photometric uncertainties were calculated from photon statistics. These uncertainties are not shown on the lightcurve plots as they were typically smaller than the data points, however, the average uncertainty for each night is given in column 12 in *Table 1*. Uncertainties due to coma, absolute calibrations, etc. were not formally estimated but are likely at least as large as the statistical uncertainties. The coherent shapes of our lightcurves suggest that such effects are minimal and can be safely ignored.

**Table 2**
Table of CCD Photometry – NOTE: Partial table. Full table available online.

| Date[a] | UT[b] | $m_R$[c] | $m_R^*$[d] | Date[a] | UT[b] | $m_R$[c] | $m_R^*$[d] | Date[a] | UT[b] | $m_R$[c] | $m_R^*$[d] | Date[a] | UT[b] | $m_R$[c] | $m_R^*$[d] |
|---|---|---|---|---|---|---|---|---|---|---|---|---|---|---|---|
| Jan 25 | 6.727 | 16.63 | 14.11 | Jan 26* | 7.664 | 16.54 | 14.12 | Jan 26* | 9.919 | 16.52 | 14.11 | Jan 26* | 12.488 | 16.88 | 14.46 |
| Jan 25 | 6.763 | 16.65 | 14.12 | Jan 26* | 7.701 | 16.53 | 14.11 | Jan 26* | 9.962 | 16.55 | 14.13 | Jan 26* | 12.525 | 16.90 | 14.48 |
| Jan 25 | 6.798 | 16.66 | 14.14 | Jan 26* | 7.739 | 16.52 | 14.10 | Jan 26* | 9.999 | 16.55 | 14.13 | Jan 26* | 12.563 | 16.89 | 14.47 |
| Jan 25 | 6.834 | 16.68 | 14.16 | Jan 26* | 7.776 | 16.51 | 14.10 | Jan 26* | 10.037 | 16.56 | 14.14 | Jan 26* | 12.606 | 16.89 | 14.48 |
| Jan 25 | 6.870 | 16.68 | 14.16 | Jan 26* | 7.814 | 16.51 | 14.09 | Jan 26* | 10.074 | 16.55 | 14.14 | Jan 26* | 12.644 | 16.89 | 14.47 |
| Jan 25 | 6.906 | 16.67 | 14.14 | Jan 26* | 7.859 | 16.51 | 14.09 | Jan 26* | 10.112 | 16.57 | 14.16 | Jan 26* | 12.681 | 16.89 | 14.47 |
| Jan 25 | 7.071 | 16.72 | 14.19 | Jan 26* | 7.896 | 16.49 | 14.07 | Jan 26* | 10.152 | 16.57 | 14.16 | Jan 26* | 12.718 | 16.88 | 14.47 |
| Jan 25 | 7.106 | 16.68 | 14.16 | Jan 26* | 7.934 | 16.50 | 14.08 | Jan 26* | 10.190 | 16.57 | 14.16 | Jan 26* | 12.756 | 16.88 | 14.46 |
| Jan 25 | 7.142 | 16.69 | 14.17 | Jan 26* | 7.971 | 16.50 | 14.08 | Jan 26* | 10.227 | 16.59 | 14.18 | Jan 26* | 12.798 | 16.88 | 14.47 |
| Jan 25 | 7.178 | 16.69 | 14.16 | Jan 26* | 8.009 | 16.49 | 14.07 | Jan 26* | 10.265 | 16.58 | 14.16 | Jan 26* | 12.835 | 16.86 | 14.44 |
| Jan 25 | 7.213 | 16.67 | 14.14 | Jan 26* | 8.050 | 16.48 | 14.06 | Jan 26* | 10.303 | 16.59 | 14.18 | Jan 26* | 12.873 | 16.85 | 14.43 |
| Jan 25 | 7.249 | 16.71 | 14.19 | Jan 26* | 8.088 | 16.48 | 14.07 | Jan 26* | 10.523 | 16.64 | 14.23 | Jan 26* | 12.910 | 16.85 | 14.43 |
| Jan 25 | 7.285 | 16.70 | 14.18 | Jan 26* | 8.125 | 16.48 | 14.06 | Jan 26* | 10.561 | 16.67 | 14.26 | Jan 26* | 12.948 | 16.83 | 14.42 |
| Jan 25 | 7.321 | 16.72 | 14.19 | Jan 26* | 8.163 | 16.49 | 14.08 | Jan 26* | 10.598 | 16.66 | 14.24 | Jan 26* | 12.990 | 16.82 | 14.40 |
| Jan 25 | 7.356 | 16.72 | 14.19 | Jan 26* | 8.200 | 16.47 | 14.05 | Jan 26* | 10.636 | 16.68 | 14.27 | Jan 26* | 13.028 | 16.82 | 14.41 |
| Jan 25 | 7.392 | 16.73 | 14.21 | Jan 26* | 8.241 | 16.48 | 14.07 | Jan 26* | 10.673 | 16.68 | 14.26 | Jan 26* | 13.065 | 16.80 | 14.39 |
| Jan 25 | 7.428 | 16.73 | 14.20 | Jan 26* | 8.279 | 16.46 | 14.05 | Jan 26* | 10.843 | 16.73 | 14.32 | Jan 26* | 13.109 | 16.80 | 14.38 |
| Jan 25 | 7.464 | 16.75 | 14.23 | Jan 26* | 8.316 | 16.46 | 14.04 | Jan 26* | 10.884 | 16.73 | 14.32 | Jan 26* | 13.146 | 16.78 | 14.37 |
| Jan 25 | 7.499 | 16.75 | 14.22 | Jan 26* | 8.354 | 16.49 | 14.07 | Jan 26* | 10.922 | 16.75 | 14.33 | Jan 26* | 13.184 | 16.76 | 14.35 |
| Jan 25 | 7.535 | 16.75 | 14.22 | Jan 26* | 8.391 | 16.46 | 14.05 | Jan 26* | 10.959 | 16.76 | 14.34 | Jan 26* | 13.221 | 16.76 | 14.34 |
| Jan 25 | 7.571 | 16.76 | 14.23 | Jan 26* | 8.431 | 16.49 | 14.07 | Jan 26* | 10.997 | 16.75 | 14.34 | Jan 26* | 13.258 | 16.74 | 14.32 |
| Jan 25 | 7.606 | 16.79 | 14.26 | Jan 26* | 8.468 | 16.45 | 14.04 | Jan 26* | 11.034 | 16.78 | 14.36 | Jan 26* | 13.298 | 16.72 | 14.30 |
| Jan 25 | 7.642 | 16.71 | 14.19 | Jan 26* | 8.506 | 16.47 | 14.06 | Jan 26* | 11.077 | 16.79 | 14.37 | Jan 26* | 13.336 | 16.70 | 14.29 |
| Jan 25 | 7.678 | 16.76 | 14.24 | Jan 26* | 8.544 | 16.46 | 14.04 | Jan 26* | 11.114 | 16.81 | 14.39 | Jan 26* | 13.373 | 16.68 | 14.26 |
| Jan 25 | 7.713 | 16.79 | 14.27 | Jan 26* | 8.581 | 16.45 | 14.04 | Jan 26* | 11.152 | 16.80 | 14.38 | Jan 26* | 13.411 | 16.68 | 14.26 |
| Jan 25 | 7.749 | 16.80 | 14.28 | Jan 26* | 8.622 | 16.47 | 14.05 | Jan 26* | 11.189 | 16.83 | 14.41 | Jan 26* | 13.448 | 16.67 | 14.25 |
| Jan 25 | 7.785 | 16.82 | 14.29 | Jan 26* | 8.659 | 16.48 | 14.06 | Jan 26* | 11.227 | 16.81 | 14.39 | Jan 26* | 13.491 | 16.67 | 14.25 |
| Jan 25 | 7.821 | 16.79 | 14.27 | Jan 26* | 8.697 | 16.48 | 14.06 | Jan 26* | 11.267 | 16.83 | 14.41 | Jan 26* | 13.528 | 16.66 | 14.24 |
| Jan 25 | 7.856 | 16.81 | 14.28 | Jan 26* | 8.734 | 16.47 | 14.06 | Jan 26* | 11.304 | 16.83 | 14.41 | Jan 26* | 13.566 | 16.66 | 14.25 |
| Jan 25 | 7.951 | 16.84 | 14.32 | Jan 26* | 8.772 | 16.46 | 14.04 | Jan 26* | 11.342 | 16.84 | 14.42 | Jan 26* | 13.607 | 16.62 | 14.21 |
| Jan 25 | 7.986 | 16.88 | 14.36 | Jan 26* | 8.814 | 16.46 | 14.04 | Jan 26* | 11.379 | 16.84 | 14.42 | Jan 26* | 13.644 | 16.61 | 14.19 |
| Jan 25 | 8.022 | 16.89 | 14.36 | Jan 26* | 8.851 | 16.45 | 14.04 | Jan 26* | 11.417 | 16.82 | 14.41 | Jan 26 | 6.853 | 16.62 | 14.21 |
| Jan 25 | 8.058 | 16.91 | 14.38 | Jan 26* | 8.889 | 16.46 | 14.04 | Jan 26* | 11.459 | 16.85 | 14.44 | Jan 26 | 6.889 | 16.60 | 14.19 |
| Jan 25 | 8.094 | 16.95 | 14.42 | Jan 26* | 8.926 | 16.46 | 14.05 | Jan 26* | 11.496 | 16.85 | 14.43 | Jan 26 | 6.925 | 16.63 | 14.22 |
| Jan 25 | 8.129 | 16.93 | 14.40 | Jan 26* | 8.964 | 16.47 | 14.05 | Jan 26* | 11.534 | 16.86 | 14.45 | Jan 26 | 6.961 | 16.62 | 14.20 |
| Jan 25 | 8.165 | 16.96 | 14.43 | Jan 26* | 9.004 | 16.45 | 14.03 | Jan 26* | 11.571 | 16.86 | 14.45 | Jan 26 | 6.996 | 16.62 | 14.21 |
| Jan 26* | 6.557 | 16.68 | 14.27 | Jan 26* | 9.042 | 16.46 | 14.04 | Jan 26* | 11.608 | 16.87 | 14.45 | Jan 26 | 7.032 | 16.60 | 14.19 |
| Jan 26* | 6.612 | 16.67 | 14.26 | Jan 26* | 9.079 | 16.45 | 14.03 | Jan 26* | 11.650 | 16.87 | 14.45 | Jan 26 | 7.067 | 16.62 | 14.20 |
| Jan 26* | 6.666 | 16.68 | 14.26 | Jan 26* | 9.117 | 16.46 | 14.04 | Jan 26* | 11.687 | 16.86 | 14.45 | Jan 26 | 7.103 | 16.55 | 14.14 |
| Jan 26* | 6.720 | 16.66 | 14.24 | Jan 26* | 9.154 | 16.47 | 14.05 | Jan 26* | 11.725 | 16.88 | 14.46 | Jan 26 | 7.139 | 16.55 | 14.14 |
| Jan 26* | 6.778 | 16.65 | 14.23 | Jan 26* | 9.196 | 16.46 | 14.04 | Jan 26* | 11.762 | 16.87 | 14.45 | Jan 26 | 7.175 | 16.58 | 14.17 |
| Jan 26* | 6.832 | 16.63 | 14.22 | Jan 26* | 9.233 | 16.47 | 14.05 | Jan 26* | 11.800 | 16.88 | 14.46 | Jan 26 | 7.284 | 16.55 | 14.14 |
| Jan 26* | 6.887 | 16.65 | 14.23 | Jan 26* | 9.271 | 16.46 | 14.04 | Jan 26* | 11.839 | 16.85 | 14.43 | Jan 26 | 7.320 | 16.57 | 14.15 |
| Jan 26* | 6.944 | 16.62 | 14.20 | Jan 26* | 9.308 | 16.47 | 14.05 | Jan 26* | 11.877 | 16.85 | 14.44 | Jan 26 | 7.356 | 16.56 | 14.14 |
| Jan 26* | 6.998 | 16.61 | 14.19 | Jan 26* | 9.346 | 16.47 | 14.05 | Jan 26* | 11.914 | 16.86 | 14.44 | Jan 26 | 7.391 | 16.55 | 14.13 |
| Jan 26* | 7.052 | 16.60 | 14.18 | Jan 26* | 9.387 | 16.48 | 14.06 | Jan 26* | 11.952 | 16.85 | 14.43 | Jan 26 | 7.427 | 16.57 | 14.16 |
| Jan 26* | 7.110 | 16.60 | 14.18 | Jan 26* | 9.424 | 16.47 | 14.06 | Jan 26* | 11.989 | 16.88 | 14.47 | Jan 26 | 7.463 | 16.54 | 14.13 |
| Jan 26* | 7.164 | 16.58 | 14.17 | Jan 26* | 9.462 | 16.48 | 14.06 | Jan 26* | 12.030 | 16.87 | 14.45 | Jan 26 | 7.498 | 16.53 | 14.11 |
| Jan 26* | 7.218 | 16.59 | 14.16 | Jan 26* | 9.499 | 16.48 | 14.07 | Jan 26* | 12.067 | 16.88 | 14.47 | Jan 26 | 7.534 | 16.52 | 14.11 |
| Jan 26* | 7.281 | 16.59 | 14.17 | Jan 26* | 9.537 | 16.48 | 14.07 | Jan 26* | 12.105 | 16.87 | 14.45 | Jan 26 | 7.570 | 16.51 | 14.10 |
| Jan 26* | 7.318 | 16.58 | 14.16 | Jan 26* | 9.579 | 16.49 | 14.08 | Jan 26* | 12.142 | 16.87 | 14.45 | Jan 26 | 7.605 | 16.52 | 14.11 |
| Jan 26* | 7.356 | 16.57 | 14.15 | Jan 26* | 9.616 | 16.51 | 14.09 | Jan 26* | 12.180 | 16.87 | 14.45 | Jan 26 | 7.641 | 16.52 | 14.11 |
| Jan 26* | 7.393 | 16.56 | 14.14 | Jan 26* | 9.654 | 16.51 | 14.09 | Jan 26* | 12.221 | 16.89 | 14.47 | Jan 26 | 7.677 | 16.50 | 14.09 |
| Jan 26* | 7.431 | 16.55 | 14.13 | Jan 26* | 9.691 | 16.51 | 14.09 | Jan 26* | 12.258 | 16.86 | 14.44 | Jan 26 | 7.712 | 16.50 | 14.09 |
| Jan 26* | 7.473 | 16.56 | 14.14 | Jan 26* | 9.729 | 16.50 | 14.09 | Jan 26* | 12.296 | 16.88 | 14.46 | Jan 26 | 7.748 | 16.51 | 14.10 |
| Jan 26* | 7.510 | 16.54 | 14.10 | Jan 26* | 9.769 | 16.52 | 14.10 | Jan 26* | 12.333 | 16.87 | 14.45 | Jan 26 | 7.784 | 16.47 | 14.06 |
| Jan 26* | 7.547 | 16.53 | 14.12 | Jan 26* | 9.807 | 16.52 | 14.11 | Jan 26* | 12.370 | 16.85 | 14.43 | Jan 26 | 7.819 | 16.50 | 14.08 |
| Jan 26* | 7.585 | 16.54 | 14.12 | Jan 26* | 9.844 | 16.53 | 14.12 | Jan 26* | 12.413 | 16.87 | 14.45 | Jan 26 | 7.855 | 16.52 | 14.11 |
| Jan 26* | 7.622 | 16.55 | 14.13 | Jan 26* | 9.882 | 16.53 | 14.11 | Jan 26* | 12.450 | 16.88 | 14.46 | Jan 26 | 7.891 | 16.47 | 14.06 |

**Notes.**
[a] UT date of observations in 2012. Data acquired with the 1.1 m telescope are denoted with an *; all other data were acquired with the 0.8 m telescope
[b] UT at midpoint of the exposure (uncorrected for light travel time).
[c] Observed $R$-band magnitude (after applying absolute calibrations, extinction corrections, and comparison star corrections.
[d] $m_R$ (1,1,0) corrected by $\Delta m_2$ (given in *Table 1*)so that all nights have the same peak magnitude.

## 3. LIGHTCURVE ANALYSIS AND INTERPRETATION

### 3.1. Rotation Period of the 2011/12 Apparition

The combined thermal IR and optical data from Millis et al. (1988) showed the shape of the comet to be approximately that of a triaxial ellipsoid, and therefore we expected a double-peaked lightcurve. In order to derive the period of the brightness variation, and thus the rotation period of the nucleus, we superimposed all the lightcurves from different nights with the data phased to a 'trial' period and zero phase set as perihelion (2011 Oct 19.1). This was possible as the geometry of the system did not change considerably throughout the apparition (see *Section 3.2*). By adjusting the trial period with a slider in Python, we could easily scan through potential rotation periods in real time and make rapid 'better-or-worse' comparisons. Whilst iterating through the different periods we looked for alignment of the peaks and troughs of the lightcurves in order to determine the optimal period as well as the period for which the data were first clearly out of phase. An example of this is shown in *Figure 4* where the data are phased to 13.44 hr, 13.45 hr and 13.46 hr. From this plot one can clearly see that 13.45 hr is in phase while 13.44 hr is too short and 13.46 hr is too long. Based on phasing the data at smaller steps of 0.001 hr, the uncertainty is estimated to be 0.005 hr. We therefore conclude a rotation period of 13.450 ± 0.005 hr for the combined data. The same process was carried out for the individual months as well as for combined adjacent months (as shown in *Figure 5* and tabulated in *Table 3*). Due to the shorter time intervals, the individual months yielded larger uncertainties than the combined months, and uncertainties increased during the apparition due to deteriorating signal-to-noise as well as shorter nightly observing windows in April and May. Phasing of data from single months revealed the shortest rotation period to be in January, with 13.45 ± 0.03 hr, and the longest in April, with 13.47 ± 0.04 hr. Combined adjacent months showed values ranging from 13.450 ± 0.005 hr in January-February to 13.458 ± 0.010 hr in April-May (see *Table 3*). These numbers hint at a small increase with time (which we will revisit below), but are consistent within the uncertainties.

**Table 3**
Summary of the determined rotation periods from three different methods and amplitude of the light curves in 2012.

| Period | Inspection | PDM | L-S | Amplitude |
|---|---|---|---|---|
| **All** | **13.450 ± 0.005** | **13.452** | **13.452** | |
| Jan | 13.45 ± 0.03 | 13.470 | 13.466 | 0.50 ± 0.05 |
| Feb | 13.45 ± 0.04 | 13.459 | 13.462 | 0.45 ± 0.1 |
| Mar | 13.46 ± 0.02 | 13.496 | 13.486 | 0.35 ± 0.15 |
| Apr | 13.47 ± 0.04 | 13.459 | 13.452 | 0.35 ± 0.1 |
| May | 13.45 ± 0.10 | 13.416 | 18.844 | 0.40 ± 0.2 |
| Jan-Feb | 13.450 ± 0.003 | 13.468 | 13.452 | |
| Feb-Mar | 13.450 ± 0.005 | 13.451 | 13.450 | |
| Mar-Apr | 13.453 ± 0.007 | 13.451 | 13.456 | |
| Apr-May | 13.458 ± 0.010 | 13.458 | 13.460 | |

A further search for periodicity was carried out using phase dispersion minimization (PDM; Stellingwerf 1978) and Lomb-Scargle (L-S; Lomb 1976, Scargle 1982) algorithms in Python 3.0. The former is a popular method often used to analyze non-sinusoidal lightcurves that have poor time coverage as it does not require uniformly sampled data. The method phases the data according to an assumed period before dividing the data into a series of bins. The individual variances of each bin are combined and compared to the overall variance of the dataset. This process is carried out for a range of trial periods. For a true period, this ratio will yield a small value θ and the phase dispersion minimization plot will reach a local minimum. The blue line in *Figure 6* illustrates the PDM for all of our 2012 data. The L-S technique, on the other hand, is similar to Discrete Fourier Transform (DFT), in that it transforms the data from the time domain to the frequency domain. However, whilst DFT usually requires evenly sampled data points, L-S does not.

Both PDM and L-S agreed on an optimal double-peaked period of 13.452 hr for the full data set. The L-S algorithm is optimized to only find the single-peaked solution, while PDM returned an equally likely single- or double-peaked solution. The L-S single-peaked result was doubled to determine the double-peaked solution since Millis et al. (1988) showed that the double-peaked solution yields the true rotation period. The doubling of the single-peaked answer will have yielded some error due to the differences in shapes of the two peaks.

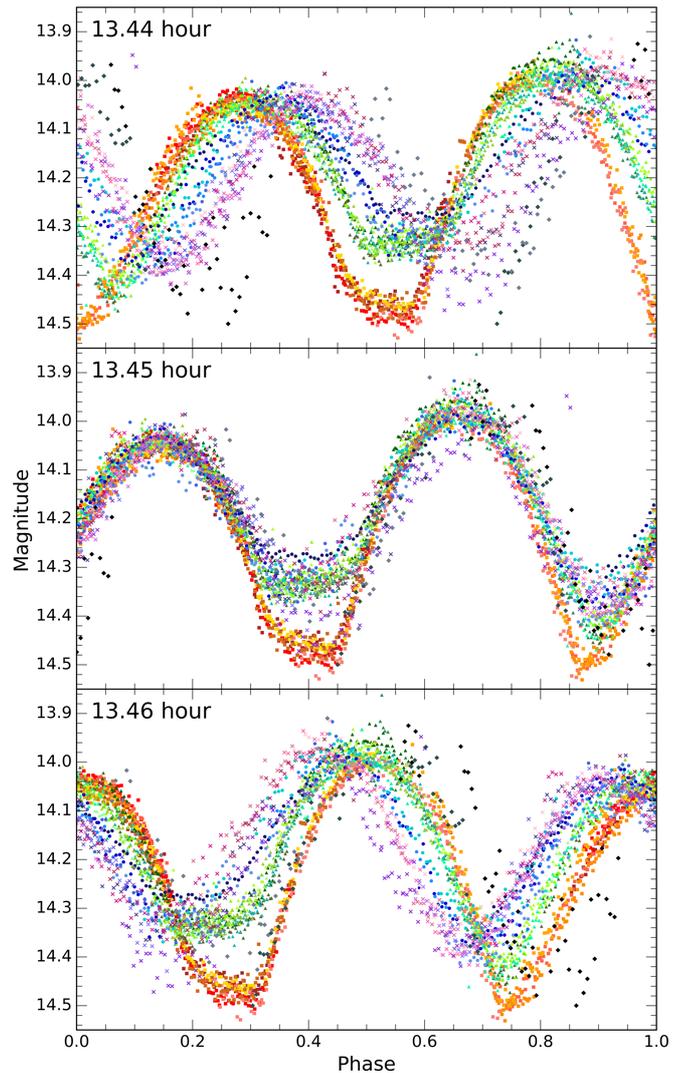

**Figure 4.** Our reduced 2012 data ($m_R$*) phased to 13.44 hr (top panel), 13.45 hr (middle panel) and 13.46 hr (bottom panel). By iterating though the different periods, we found a best period of 13.450 ± 0.005 hr. Zero phase was set at perihelion (2011 October 19.1). Lightcurves are aligned to the peak brightness, as these were more consistent throughout the apparition than the troughs. The nightly points are as given in *Figure 1*.

Values for various subsets of the data are summarized in *Table 3*.

The uncertainties associated with both PDM and L-S are indeterminate. *Figure 6* shows that, even though the PDM algorithm presents a distinct lowest θ, it is uncertain how far from the absolute minimum can still be considered a viable solution. This is also the case with L-S. Conversely, with the manual phasing of the data to a number of trial periods, we were able to identify where the phasing broke down; this was greatly aided by the use of different colors for different days, as shown in *Figure 4*.

As a whole, the rotation periods obtained by inspection agree well with the values obtained through PDM and L-S, to within reasonable uncertainties. The exception to this being the values of the PDM and L-S for 2012 March and the value of the L-S for 2012 May, which are unreasonably high at 13.496 hr, 13.486 hr and 18.844 hr respectively. As illustrated in *Figure 4*, even deviations by 0.01 hr from a period of 13.45 hr result in the lightcurves to be significantly out of phase. Based on this we conclude that these solutions are incorrect.

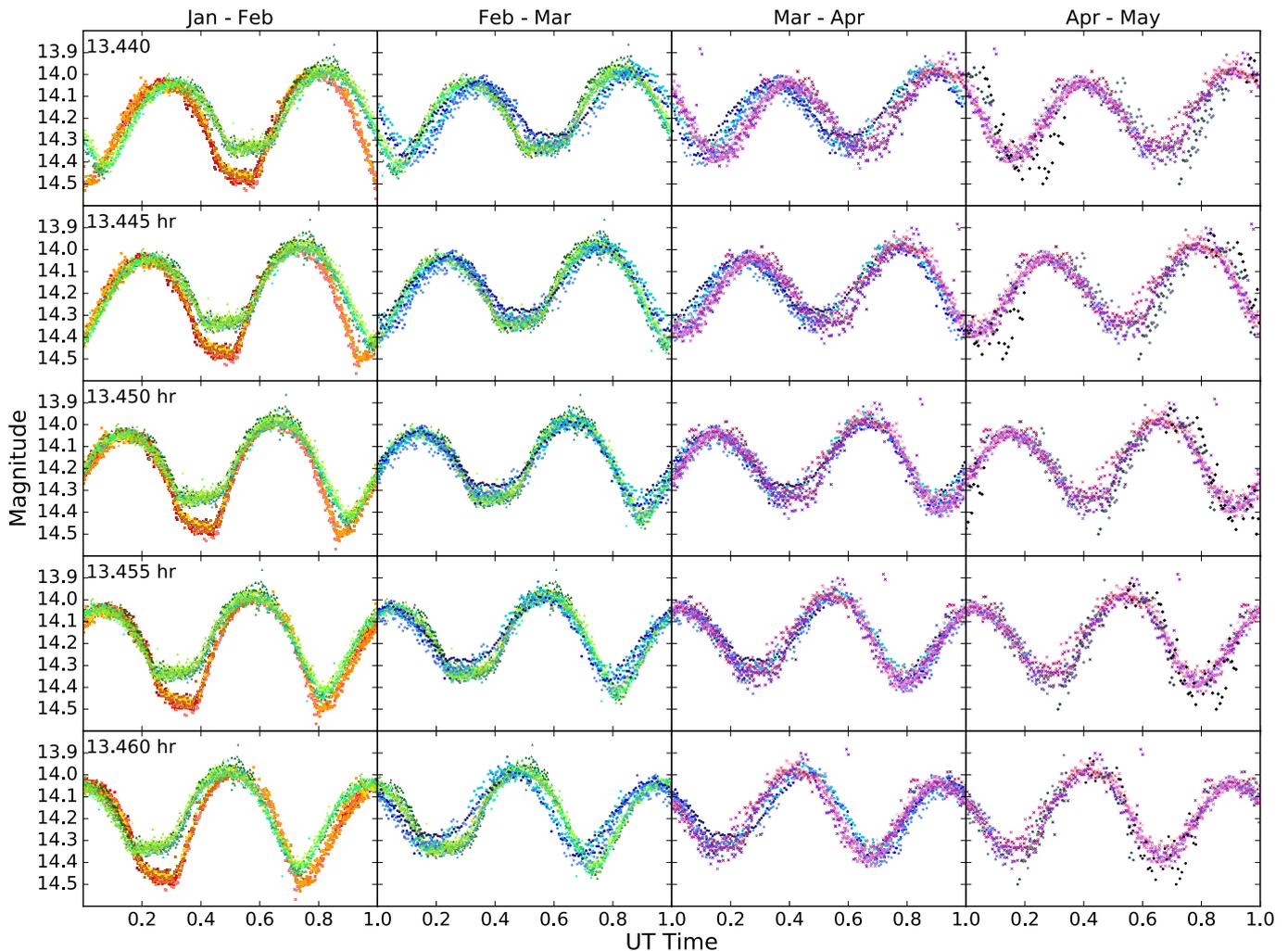

**Figure 5.** Reduced magnitude ($m_R^*$) lightcurves for pairs of months: January (orange), February (green), March (blue), April (pink) and May (black). Columns show pairs of months (as labelled above each column) while the rows show the same rotation period (labelled in the first column). The points are as given in *Figure 1*.

### 3.2. *Implications of Viewing Geometry*

The rotation period obtained through phasing of the data is the time it takes for the brightness to appear the same as viewed from Earth, known as the synodic period. Conversely, the sidereal period is the period relative to a fixed point in space, and is the 'true' rotation period. As both the Earth and the comet are moving in their orbits, the geometry of the system changes, resulting in different parts of the comet being illuminated and hence in subtle changes in the synodic period. In order to confirm that this did not affect our determined rotation periods, and in particular our comparison of results between different apparitions (*Section 3.4*), the extent of this effect was assessed.

This assessment was based, in part, on our assumption that we were viewing the comet near equator-on, e.g., the comet's rotational pole was near the plane of sky. If the comet was not being viewed near equator-on, the large amplitude observed in our lightcurves would only occur if the comet was highly elongated, and we have no reason to believe that this is the case. Furthermore, the amplitude of the lightcurves from the 1984/85 apparition (see *Section 3.4*) were found to agree with the amplitude of the lightcurves from the 2011/12 apparition. Due to the differences in viewing geometry between apparitions, the similar amplitudes suggest that we are viewing the comet at similar

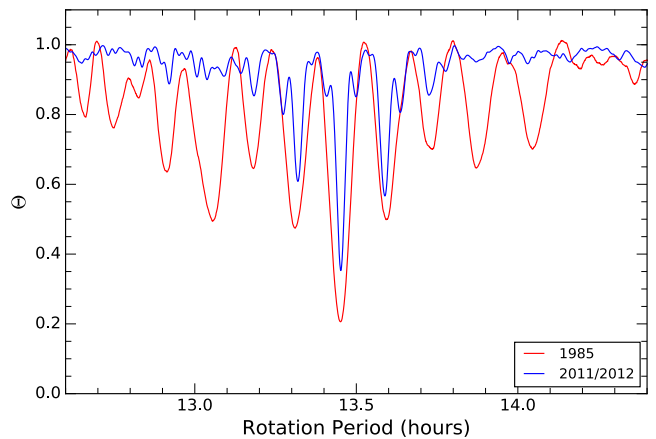

**Figure 6.** Phase Dispersion Minimization of our 2012 data (blue) compared to the combined data from the 1985 apparition (red; Millis et al. 1988, Jewitt & Meech 1985, Wisniewski et al. 1986).

sub-Earth latitudes and hence, that the rotational pole is near the plane of the sky.

As discussed above, the 2012 data showed a hint of an increase in the rotation period from January to May, although they are consistent with a constant value within the uncertainties (see *Table 3*). Without consideration of the Earth's geometric position, this is suggestive of retrograde rotation (obliquity near 180°), which would result in the sidereal period being longer than our measured synodic period. The prograde case (obliquity near 0°), would have a sidereal period shorter than our measured synodic period by a comparable amount.

For an obliquity of 180°, the offset between the synodic and sidereal periods ranges from 0.010 hr to 0.005 hr between 2012 January and May. As our uncertainties are smallest in January, we use 0.010 hr as the most likely synodic-sidereal offset, resulting in a sidereal rotation period of 13.460 hr when only the solar component is considered. During this same interval, the viewing angle from Earth changed much less, and thus the phase angle bisector (cf. Harris et al. 1984) varied by only about half of the solar component alone, implying a somewhat smaller sidereal value. Even though we have reason to believe that the pole is near the plane of the sky, there is evidence for strong seasonal effects (see *Section 4.2*). Our assumptions, however, change minimally even if the axis is intermediate, e.g., the synodic-sidereal offset is only significantly different if the pole is nearly perpendicular to the plane of the sky.

### 3.3. Lightcurve Shape

The lightcurves of the 2011/12 apparition show a clear asymmetry, with one sharp, deeper trough (near phase 0.9 in the middle panel of *Figure 4*) and one flatter, shallower trough (near phase 0.4 in the middle panel of *Figure 4*). In addition, the peak-to-peak times of all the monthly lightcurves are larger than the trough-to-trough times with the latter being approximately 10% shorter. These asymmetries, which are likely due to the shape of the nucleus deviating from that of a simple triaxial ellipsoid due to e.g., large boulders or flat areas, reduce the uncertainty in the period, as they highlight a clear correct phase and eliminate solutions that are a half phase off. The sudden change in lightcurve shape from February to March, with the sharp trough disappearing, further suggests deviations from the triaxial ellipsoid (e.g., Durech et al. 2011). These distinct features in the lightcurve can also be seen in the 1985 data (see *Section 3.4*).

As seen in *Figure 5* and tabulated in *Table 3*, the amplitudes of the lightcurves vary by approximately 0.15 magnitudes between 2012 January and May. This could be due to a change in orientation of the comet relative to us, resulting in a change in the apparent cross section. Furthermore, the coma suppresses the nucleus contribution, resulting in a decrease in the amplitude of the lightcurve. Removal of the coma would increase the amplitude by around 10 – 20%; however, this would also greatly increase the uncertainties, as previously discussed. In addition to the effects of the coma, the position angles of the Sun and the solar phase angles (columns 7 and 8 of *Table 1*) imply that the tail is highly projected, resulting in a large amount of tail remaining in the photometric aperture. The effect of this was not formally assessed. The uncertainty in the amplitude steadily increased between January and May as the comet became fainter and thus the signal-to-noise got worse.

The minimum axis ratio of the nucleus can be calculated from the observed amplitude of the lightcurve using the equation (e.g., Mueller & Ferrin 1996):

$$10^{-0.4(m_{min}-m_{max})} \geq \frac{b}{a} \qquad [2]$$

where $m$ is the magnitude and $a$ and $b$ are the semi-major and semi-minor axes respectively. The peak-to-trough variation of the lightcurves ranged from 0.35 to 0.50 (*Table 3*), corresponding to minimum axial ratios of 1.38 and 1.63. This is in agreement with the axial ratio of 1.6 that was obtained by Millis et al. (1988) by averaging optical and infrared amplitudes and confirms that coma contamination was minimal.

Although we have elected not to remove the coma contamination for our determination of the rotation period, a first-order removal yields a plausible estimate of the nucleus size. For example, on February 26 the middle of the lightcurve was at an apparent magnitude of $m_R$=16.38 and we estimated 15% of the aperture flux came from coma contamination. Removal of the coma yields a nucleus magnitude of 16.56, which can be converted to a nuclear radius by the standard methodology (e.g., Jewitt 1991). Assuming a geometric albedo of 0.028 (Millis et al. 1988) and a nucleus solar phase angle correction of 0.04 magnitudes per degree, we estimate an effective radius of 4.6 km for this night. Similar calculations throughout the apparition yield effective radii in the range 4.4-4.8 km, in excellent agreement with Kelley et al. (2017) who found an effective radius of 4.57 km using thermal modeling of mid-IR data.

### 3.4. Reanalysis of the 1985 Data

We reanalyzed the publicly available data from three independent groups collected during the favorable 1984/85 apparition. Millis et al. (1985) derived a rotation period of 13.47 ± 0.02 hr, based on their optical observations spanning six nights in late January 1985. Further optical observations of comet 49P/Arend-Rigaux were made by Jewitt & Meech (1985) on four consecutive nights between 1985 January 18 and 21. Their optical observation showed lightcurves with a single peaked period of either 9.58 ± 0.08 or 6.78 ± 0.08 hr, or a multiple of one of these. Similarly, Wisniewski et al. (1986) observed the comet on a total of eight nights (1985 January 17-21, 1985 February 15-17), derived a quadruple peaked rotation period of 27.312 hr. Based on the thermal IR data from Millis et al. (1988), as well as the asmmetry observed in the 2012 data, we eliminate the single and quadruple peaked solutions.

None of these data sets were ideal in terms of both removing aliases and obtaining precision. This was largely due to the limited amount of temporal coverage acquired by any one group, as well as the lack of knowledge of the shape of the nucleus (Jewitt & Meech 1985; Wisniewski et al. 1986). In order to improve upon their individual results, we combined the data from the three independent groups, thus significantly increasing the baseline and allowing us to eliminating potential aliases as well as increasing the overall precision. Whilst the first two papers tabulated their data, Wisniewski et al. (1986) only presented figures of their results, which were phased using their preferred rotation period. See the Appendix for details of the procedure used to extract the data that we required. Prior to phasing the data, we arbitrarily adjusted the lightcurves in order to bring them all to the same peak magnitude. The large differences in magnitudes between the different data sets were due to the lack of instrumental magnitude correction in the Jewitt & Meech (1985) data, as well as methodological differences and small changes in the geometry or the comet's activity; amplitudes also differ, consistent with each group's use of a different aperture size resulting in differing amounts of coma contamination.

The rotation period of the 1985 data was determined in the same way as described in *Section 2* for the 2012 data. We obtained a value of 13.45 ± 0.01 hr by inspection and values of 13.450 hr and 13.448 hr using PDM and L-S, respectively, with the phased results shown in *Figure 7*. Within reasonable uncertainty, these values are consistent with the rotation period of 13.47 hr reported by Millis et al. (1988).

As shown in *Section 3.2*, the offset between the synodic and sidereal periods is between 0.010 hr and 0.005 hr during the 2011/12 apparition for an obliquity of 180°. Similarly, for the same obliquity, there was an average offset of 0.012 hr during the published 1985 observations. Since the offsets are in the same direction during each apparition, the relative effect differs by a maximum of 0.007 hr, a value that lies within the uncertainties of the synodic period of either of these apparitions. This shows that it is safe to ignore the synodic-sidereal effects when intercomparing the rotation periods of the two apparitions, and we can compare the rotation periods directly. Our determined rotation period for the 1984/85 apparition agrees with the values obtained for the 2011/12 apparition within the calculated uncertainties, constraining the maximum change in rotation period to 54 s (0.015 hr). As there were four intervening perihelion passages, this is equal to a maximum change of less than 14 s per apparition.

## 4. COMA PROPERTIES

### *4.1. Unexpected Feature*

As noted previously, the $\varDelta m_2$ values for 2012 March 16-22 stand out as unusually large, suggesting the comet was brighter than expected by 0.2-0.3 mag. Nightly stacks of images during this time revealed a jet-like feature in a direction very different from the expected tail direction of older material (*Figure 8*). The jet-like feature first appears in the stacked images of March 16, where the last observed night prior to this date, February 26, showed no sign of activity (see *Figure 2*). The feature continued to grow in projected length until it separated from the nucleus around March 25. In order to better determine the point of separation, we removed individual frames with significantly worse seeing as well as multiple images on March 27 that were contaminated by a bright star passing through the feature.

Unlike a normal, e.g., sublimation-driven, jet, we believe this to have been an impulse type outburst, which has an elongated appearance due to the range of particle sizes and masses travelling radially outwards at different velocities.

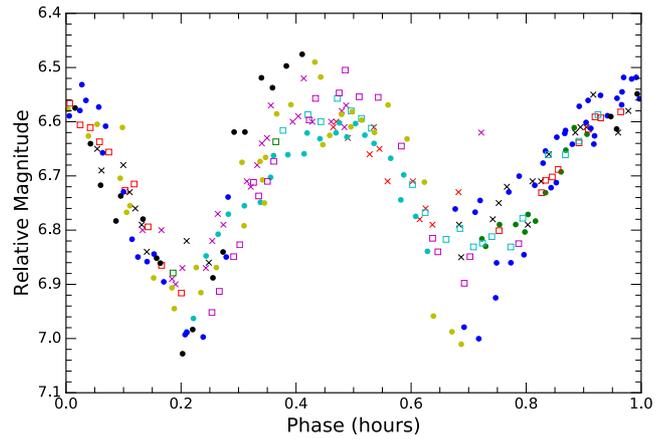

**Figure 7.** Reanalysis of the 1985 data. The open squares display the data from Millis et al. (1988), the crosses the data from Jewitt & Meech (1985) and the filled circles data from Wisniewski et al. (1986). Different colors are used for different nights by the same authors. The best rotation period is found to be 13.45 ± 0.01 hr.

Based on the relatively narrow angular width of the outburst, we hypothesize that the duration of the event was less than ~2 hours. If the event had gone on for a longer period of time, we would expect to see an increased amount of angular spreading of the feature due to the rotation of the comet (unless the jet was near-polar). Detailed modeling of the evolution of the jet's shape and extent would likely constrain aspects of the outburst such as source orientation, grain sizes, and duration but is beyond the scope of this paper; however, some properties can be derived.

In order to extrapolate back to a time of the onset of activity, the distances from the nucleus to the trailing and leading extent of the jet-like feature were measured for each night. Based on the assumption that the grains travel at a constant projected velocity, a trend line enabled us to extrapolate backwards to the point where the grains originated. This was found to be on March 15 around 18 hours UT and is presented as time zero hours in *Figure 9*. The trailing and leading particles traveled at projected velocities of 17 m s$^{-1}$ and 56 m s$^{-1}$, respectively, and the near constant velocity implies that the acceleration due to radiation pressure was primarily in our line of sight (consistent with the solar phase angle being near 15°) and thus had minimal effect on the projected velocity.

There is no evidence of a similar event in the previously analyzed apparitions; however, it cannot be ruled out that this is a seasonal effect rather than an isolated outburst. Regardless of its origin, an order of magnitude calculation of the quantity of material involved shows that it is trivial compared to the large size of the nucleus and would not have had a discernible effect on the rotation period. Based on the $\Delta m_2$ values in *Table 1*, we can crudely estimate that the cross section, $C$, of material released by the outburst was ~30% of the total nucleus cross section. This can be converted to a mass, $M$, by $M = (4/3) \times \rho \times a_{avg} \times C$ (e.g., Jewitt 2013) where $a_{avg}$ is the average particle radius (assumed to be 1 micron) and $\rho$ is the material density (assumed to be 1900 kg m$^{-3}$; Rotundi et al. 2015), yielding ~5×10$^4$ kg. For reasonable assumptions about the bulk density of Arend-Rigaux (~500 kg m$^{-3}$) and dust-to-gas ratio (~1), this mass of material can be easily explained by the excavation of a hemispherical pit <10 m in radius. This is comparable to or smaller than many pits observed on the surface of 67P/Churyumov-Gerasimenko by *Rosetta* (e.g., Sierks et al. 2015), and significantly smaller than the crater produced by the Deep Impact experiment (200 ± 20 m diameter; Schultz et al. 2013). Thus, an outburst such as this is likely unexceptional, and it should come as no surprise that it did not produce a detectable change in the rotation period.

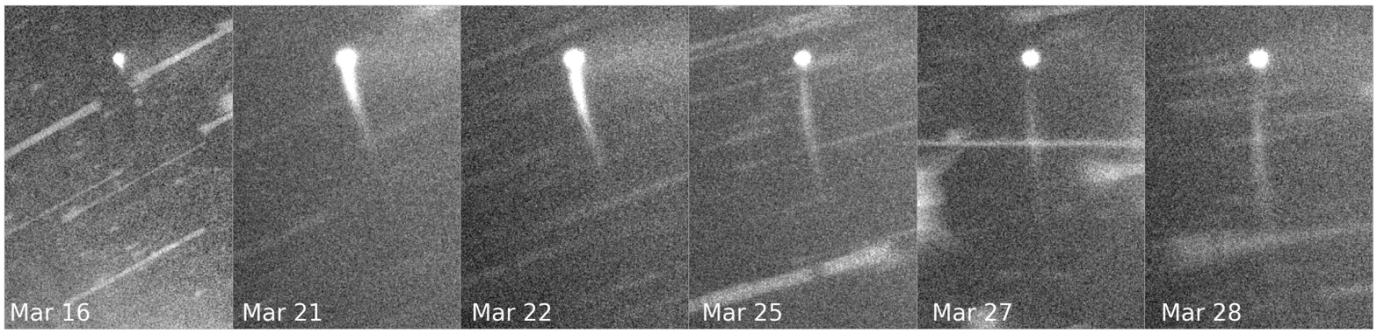

**Figure 8.** Nightly stacked and median combined images between 2012 March 16 – 28. All images present the same physical size at the comet of 112,000 km by 160,000 km and are oriented so north is up and east is to the left. The jet-like feature was first observed on 2012 March 16 and was seen throughout the remainder of the month. The expected tail direction of old material during this time was PA~290° (JPL Horizons).

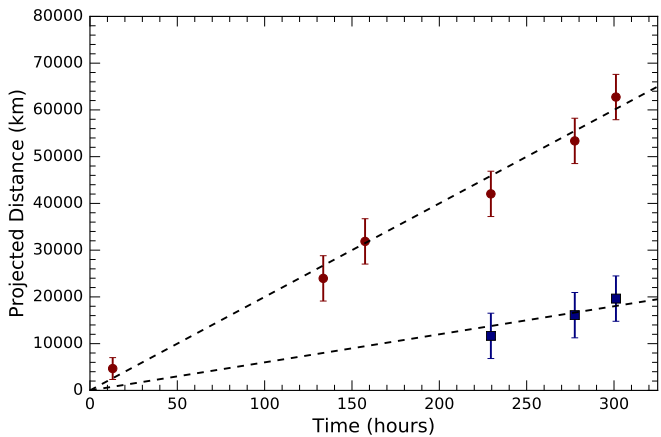

**Figure 9.** The position of the trailing (blue squares) and leading (red circles) extent of the jet-like feature relative to the nucleus in km, where 0 hours is defined as the most likely time of the onset of activity on 2012 March 15 about 18 hours UT. The lines of best fits suggest minimum particles velocities of 17 m s$^{-1}$ and 56 m s$^{-1}$ for the trailing and leading extent of the outburst respectively. Average uncertainties of 6 pixels for the 1.1 m telescope and 12 pixels for the 0.8 m telescope were determined based on seeing and on the ease of determining the edges of the jet.

### 4.2. Gas and Dust Production Rates

Overall, comet 49P/Arend-Rigaux was simply too faint for us to obtain our standard narrowband photometric measurements of the coma during the 2011/12 apparition. However, we were able to obtain data during its 2004/05 apparition, when it was somewhat brighter but only available for one or two sets per night due to the short observing window from our northern hemisphere location and having competing targets. Here we present these data, along with a reanalysis of similar observations obtained in 1984/85 by Millis et al. (1988), so that both data sets utilize the same reduction parameters; note that in particular the Haser scalelengths and daughter lifetimes used by us to derive gas production rates changed in the decade following the Millis et al. (1988) paper.

Using our now standard observing and reduction procedures (cf. A'Hearn et al. 1995; Schleicher & Bair 2011), observations at both apparitions were obtained with photoelectric photometers using narrowband comet filters (the IHW set in 1984/85 and the HB set in 2004/05; cf. Osborn et al. 1990, Farnham et al. 2000). Reduced fluxes, aperture abundances, and production rates were computed for each gas species -- OH, NH, CN, C$_3$, and C$_2$. We also compute abundance ratios, water production rates, the effective active area on the surface of the nucleus required to produce the water based on a standard vaporization model, and the active fraction based on the surface area of the nucleus. For the dust, the fluxes and the now standard proxy for dust production, $Af\rho$, (A'Hearn et al. 1984b) are determined from the continuum measurements (see *Tables 4, 5* and *6*). Because of the wide range of solar phase angles, particularly in 1984/85, phase adjustments were made to yield $A(0°)f\rho$. Furthermore, due to evidence for trends in $Af\rho$ with aperture size (with a very wide range of aperture sizes) we apply an aperture adjustment.

In *Figure 10*, we plot the log of the production rates for each species with respect to time from perihelion. In spite of the fact that the temporal coverage at each apparition is sparse, it is evident that there is a significant pre-/post-perihelion asymmetry, with production rates as much as 50% to 100% greater during comet 49P/Arend-Rigaux's approach to the Sun. Less certain is the time of peak production because of differences between the two apparitions; we estimate that peak production occurred near $\Delta T$ ~ -20 days. Both properties imply a seasonal effect due to a changing sub-solar latitude and one or more active source regions, rather than uniform leakage of gas from the entire surface. However, as indicated in *Section 3.3*, the obliquity of the pole cannot be too large or we should have seen a significant change in the lightcurve amplitudes as a function of viewing geometry.

**Table 4**
Photometry Observing Circumstances and Fluorescence Efficiencies for Comet 49P/Arend-Rigaux.

| UT Date | $\Delta T$ (day) | $r_H$ (AU) | $\Delta$ (AU) | Solar Phase Angle (°) | Phase Adj. log $A(0°)f\rho^a$ | $\dot{r}_H$ (km s$^{-1}$) | log $L/N^b$ (erg s$^{-1}$ molecule$^{-1}$) OH | NH | CN | Telescope |
|---|---|---|---|---|---|---|---|---|---|---|
| 1984 Oct 26.51 | −35.53 | 1.499 | 0.918 | 40.2 | +0.44 | −5.0 | −15.070 | −13.457 | −12.796 | 1.8 m |
| 1984 Dec 21.53 | +20.49 | 1.463 | 0.595 | 28.7 | +0.37 | +2.9 | −14.983 | −13.439 | −12.764 | 1.8 m |
| 1985 Jan 26.48 | +56.44 | 1.571 | 0.590 | 5.0 | +0.09 | +7.2 | −14.917 | −13.416 | −12.728 | 2.2 m |
| 1985 Jan 27.35 | +57.31 | 1.575 | 0.593 | 4.7 | +0.08 | +7.3 | −14.921 | −13.417 | −12.730 | 2.2 m |
| 1985 Jan 28.43 | +58.39 | 1.579 | 0.597 | 4.6 | +0.08 | +7.4 | −14.921 | −13.419 | −12.732 | 2.2 m |
| 1985 Feb 15.32 | +76.28 | 1.664 | 0.705 | 12.6 | +0.20 | +8.9 | −14.951 | −13.474 | −12.777 | 1.8 m |
| 2004 Dec 10.18 | −75.88 | 1.614 | 1.106 | 36.8 | +0.43 | −9.8 | −15.040 | −13.514 | −12.876 | 1.1 m |
| 2005 Mar 8.15 | +12.09 | 1.375 | 1.320 | 43.2 | +0.45 | +1.9 | −15.038 | −13.415 | −12.772 | 1.1 m |
| 2005 Mar 10.18 | +14.12 | 1.377 | 1.330 | 43.0 | +0.45 | +2.3 | −15.013 | −13.404 | −12.747 | 1.1 m |
| 2005 Apr 6.19 | +41.11 | 1.445 | 1.506 | 39.6 | +0.44 | +6.2 | −14.889 | −13.355 | −12.654 | 1.1 m |

**Notes.**
$^a$ Adjustment to 0° **solar** phase angle to $A(\theta)f\rho$ values based on assumed phase function (see text).
$^b$ Fluorescence efficiencies are for $r_H$ = 1 AU, and are scaled by $r_H^{-2}$ in the reductions.

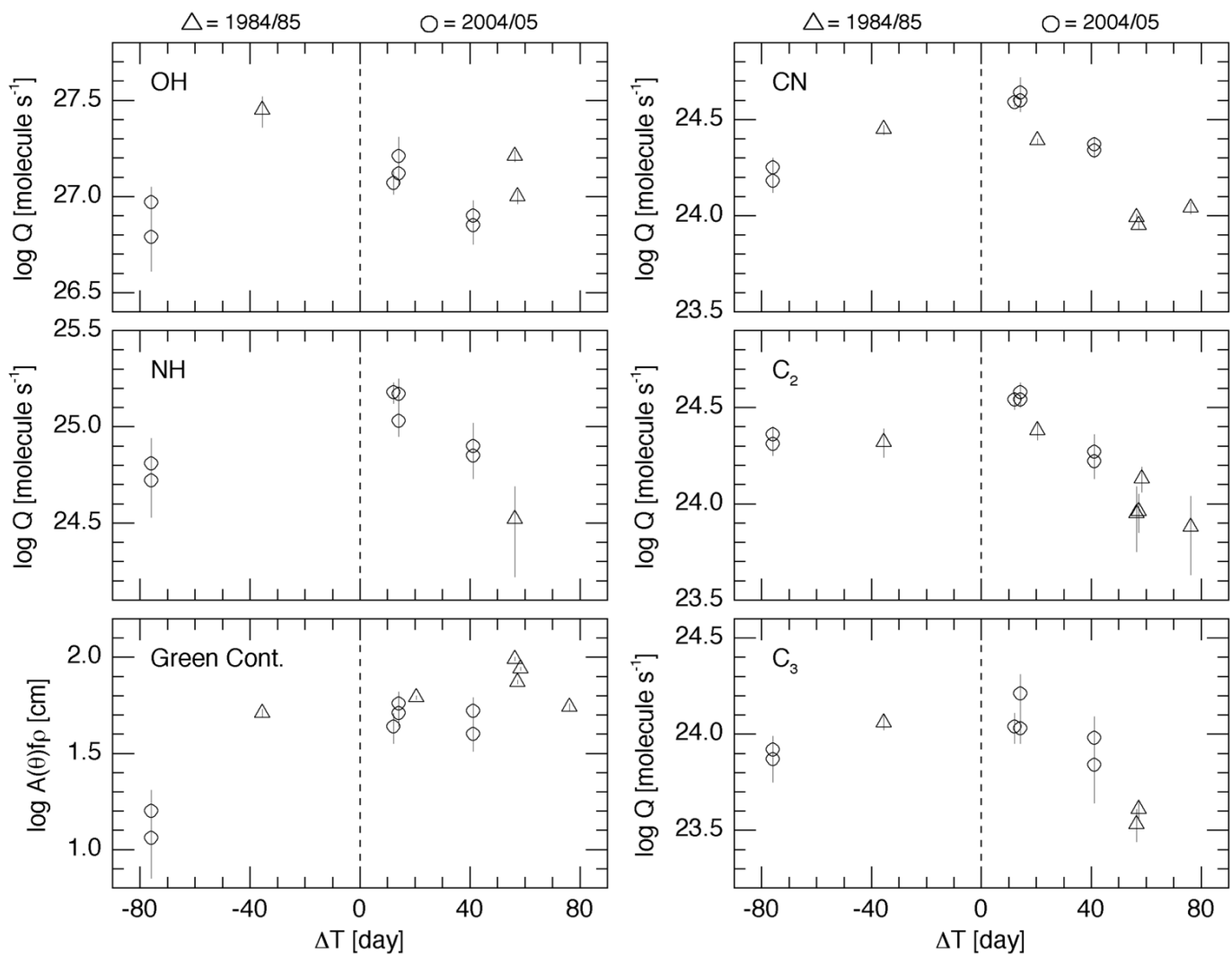

**Figure 10.** Log of the production rates for each observed molecular species and $A(\theta)f\rho$ for the green continuum plotted as a function of time from perihelion. Data points from the 1984/85 apparition are shown as triangles while those from 2004/05 are shown as circles. Even with so few pre-perihelion points, it is evident that each of the gas species exhibit a significant seasonal effect with production rates substantially lower following perihelion, indicative of a source region moving from summer towards winter. The opposite behavior exhibited by the dust is entirely an artifact, primarily due to phase effects and secondarily due to a trend with aperture size (see text and *Figure 11*). There is also possible evidence for a long-term secular change, but with the minor species increasing between 1984/85 and 2004/05 while OH decreases.

Inter-comparison of the 1984/85 and 2004/05 data reveals a surprise: CN and $C_3$ clearly imply higher values at the later apparition, $C_2$ and NH are less certain but also consistent with this, but OH and the dust exhibit the opposite long-term secular trend. As discussed later, the apparent dust behavior is primarily an artifact due to phase effects and aperture trends, but the OH is a puzzle. In our photometric database (Schleicher & Bair 2016) we have many examples of the OH having differing amounts of asymmetry or differing $r_H$-dependencies from the minor species. Comet 49P/Arend-Rigaux is the first case where OH and the minor species exhibit secular changes in opposite directions, usually indicative of at least two source regions having different compositions and a possible precession of the pole. However, significant precession seems highly unlikely due to the lack of change in the rotation period, thus little or no evidence of torquing, coupled with the small outgassing rates and the large nucleus size. Also, the OH secular change is large; we estimate about 2 times greater in 2004/05 than in 1984/85. We therefore tentatively conclude that a change in the relative outgassing rates between two source regions (having different relative abundance of minor species vs water) is *not* due to solar insolation but rather changes to the source regions themselves.

In spite of the secular variations seen in the relative abundances, comet 49P/Arend-Rigaux remains in the 'typical' compositional group throughout (A'Hearn et al. 1995; Schleicher & Bair 2016). Water production rates, based on OH and listed in *Table 6*, imply an effective active area that ranges from 0.53 km$^2$ to 2.27 km$^2$ over the entire dataset, using a vaporization model by A'Hearn[1] based on the work of Cowan & A'Hearn (1979) for a pole-on, rapidly rotating nucleus. The overall mean value is 1.00 km$^2$ while the median value is smaller at 0.88 km$^2$. When combined with the effective radius given earlier of 4.57 km (Kelley et al. 2017), it yields an active fraction of 0.38% (mean) or 0.34% (median). In the context of our entire photometric database, comet 49P/Arend-Rigaux thus has the fifth lowest active fraction. The most extreme is recently investigated 209P/LINEAR at ~0.024% (Schleicher & Knight 2016), followed by 28P/Neujmin 1 at ~0.05%, P/LONEOS (2001 OG10) at ~0.06%, and P/Siding Spring 3 (2006 HR30) at

---

[1] http://www.astro.umd.edu/~ma/evap/

**Table 5**
Photometric Fluxes and Aperture Abundances for Comet 49P/Arend-Rigaux.

| | Aperture | | log Emission Band Flux (erg cm$^{-2}$ s$^{-1}$) | | | | | log Continuum Flux[a] (erg cm$^{-2}$ s$^{-1}$ Å$^{-1}$) | | | log M($\rho$) (molecule) | | | | |
|---|---|---|---|---|---|---|---|---|---|---|---|---|---|---|---|
| UT Date | Size (arcsec) | log $\rho$ (km) | OH | NH | CN | C$_3$ | C$_2$ | UV | Blue | Green | OH | NH | CN | C$_3$ | C$_2$ |
| 1984 Oct 26.51 | 28.5 | 3.98 | –11.59 | — | –12.12 | –11.78 | –12.36 | –14.74 | — | –14.26 | 30.85 | — | 28.05 | 27.95 | 27.72 |
| 1984 Dec 21.53 | 28.5 | 3.79 | — | — | –12.07 | — | –12.20 | –14.21 | — | –13.97 | — | — | 27.69 | — | 27.48 |
| 1985 Jan 26.48 | 20.0 | 3.63 | –11.92 | –13.37 | –12.73 | –12.46 | –13.00 | –14.28 | — | –13.98 | 29.99 | 27.04 | 26.98 | 26.92 | 26.73 |
| 1985 Jan 27.35 | 40.3 | 3.94 | –11.61 | — | –12.26 | –11.95 | –12.47 | –14.13 | — | –13.80 | 30.31 | — | 27.46 | 27.44 | 27.26 |
| 1985 Jan 28.43 | 28.5 | 3.79 | — | — | — | — | –12.56 | — | — | –13.89 | — | — | — | — | 27.18 |
| 1985 Feb 15.32 | 28.5 | 3.86 | — | — | –12.53 | — | –12.91 | –14.46 | — | –14.21 | — | — | 27.39 | — | 27.02 |
| 2004 Dec 10.17 | 62.4 | 4.40 | –11.74 | –12.43 | –12.02 | –11.67 | –11.96 | –14.72 | –14.54 | –14.59 | 30.83 | 28.62 | 28.39 | 28.28 | 28.34 |
| 2004 Dec 10.20 | 97.2 | 4.59 | –11.27 | –12.20 | –11.67 | –11.56 | –11.64 | –14.68 | –14.48 | –14.53 | 31.31 | 28.85 | 28.74 | 28.39 | 28.66 |
| 2005 Mar 8.15 | 62.4 | 4.48 | –11.42 | –11.90 | –11.48 | –11.53 | –11.58 | –14.30 | –13.97 | –14.09 | 31.31 | 29.20 | 28.98 | 28.44 | 28.73 |
| 2005 Mar 10.17 | 62.4 | 4.48 | –11.35 | –12.05 | –11.46 | –11.54 | –11.59 | –14.14 | –14.15 | –14.02 | 31.36 | 29.05 | 28.99 | 28.43 | 28.74 |
| 2005 Mar 10.20 | 38.5 | 4.27 | –11.58 | –12.24 | –11.71 | –11.54 | –11.85 | –15.61 | –14.12 | –14.19 | 31.13 | 28.86 | 28.73 | 28.44 | 28.47 |
| 2005 Apr 6.17 | 38.5 | 4.32 | –11.87 | –12.50 | –11.94 | –12.01 | –12.25 | –15.11 | –14.57 | –14.32 | 30.83 | 28.66 | 28.52 | 28.12 | 28.23 |
| 2005 Apr 6.19 | 62.4 | 4.53 | –11.50 | –12.22 | –11.67 | –11.69 | –11.99 | –14.80 | –14.11 | –14.23 | 31.20 | 28.94 | 28.79 | 28.43 | 28.48 |

**Notes.**
[a] Continuum filter wavelengths: UV (1984/85) = 3650 Å, UV (2004/05) = 3445 Å; blue (2004/05) = 4450 Å; green (1984/85) = 4845 Å, green (2004/05) = 5260 Å.

~0.13% (Schleicher & Bair 2016). Interestingly, while the first two are Jupiter-family objects, the latter two are both in the Halley-type dynamical class and presumed to originate from the Oort Cloud rather than the Kuiper Belt, implying Oort Cloud comets can also evolve to a nearly inert state.

As noted earlier, the dust production, as given by $A(\theta)f\rho$, differs greatly from those of the minor gas species, most closely resembling the behavior of OH (see *Figure 10*). However, this perception is an artifact due to a combination of viewing circumstances, specifically solar phase angle effects, and the plate scales of the telescopes used, associated with aperture trends. In particular, while all of the 2004/05 observations were taken at a narrow range of solar phase angles (37°- 43°), only the first night in 1984 had a comparable value (40°) while on later nights the solar phase angle ranged between 5° and 28°. We therefore normalized the results to 0° solar phase angle by applying our composite phase curve (cf. Schleicher & Bair 2011 and references therein); the specific adjustment factors are listed in *Table 4*. As is evident from the table, $Af\rho$ for the largest solar phase angles are adjusted by 2.3 times more than for the smallest angles, negating the apparent increase in $Af\rho$ seen near the end of the apparition for the 1985 observations.

Comet 49P/Arend-Rigaux also exhibited a trend with aperture size in $Af\rho$, with larger apertures yielding smaller values, implying a steeper radial profile for the dust than the canonical $1/\rho$ expected for coasting and unchanging grains. This is not a surprise as few comets actual follow the $1/\rho$ curve, but the small number of cases where two apertures were measured on a given night made determining an appropriate adjustment difficult. While we might normally just note the issue but not make any adjustments, the nearly order of magnitude range in projected aperture sizes, with $\rho$ varying from 4300 km to 38,900 km, requires a nominal adjustment. Based on the trends observed, including that from the imaging in early 2012, we have normalized all log $A(0°)f\rho$ values to log $\rho$ = 4.0, using an adjustment of 0.02 in the log for each 0.10 change in log $\rho$. Thus $Af\rho$ for the largest projected radius (log $\rho$ = 4.59) increases by 31% when normalized to 10,000 km, while the smallest value (log $\rho$ = 3.63) decreases by 19%.

The resulting phase adjusted and aperture normalized dust results are presented in *Figure 11*, and it is evident that dust production most closely matches the seasonal and secular behavior exhibited by CN. The dust to gas ratio, based on the adjusted $Af\rho$ values divided by $Q$(OH), vary by nearly a factor of four across the apparitions and with time, from only slightly dustier than average for our database (Schleicher & Bair 2016) in early 1984/85 to about 4 times the average in

**Table 6**
Photometric Production Rates for Comet 49P/Arend-Rigaux.

| | $\Delta T$ | log $r_H$ | log $\rho$ | log $Q^a$ (molecule s$^{-1}$) | | | | | log $A(\theta)f\rho^{a,b}$ (cm) | | | log $Q$ |
|---|---|---|---|---|---|---|---|---|---|---|---|---|
| UT Date | (day) | (AU) | (km) | OH | NH | CN | C$_3$ | C$_2$ | UV | Blue | Green | H$_2$O |
| 1984 Oct 26.51 | –35.53 | 0.176 | 3.98 | 27.45 .07 | — | 24.45 .02 | 24.06 .04 | 24.32 .07 | 1.46 .08 | — | 1.71 .02 | 27.50 |
| 1984 Dec 21.53 | +20.49 | 0.165 | 3.79 | — | — | 24.39 .01 | — | 24.38 .04 | 1.78 .04 | — | 1.79 .01 | — |
| 1985 Jan 26.48 | +56.44 | 0.196 | 3.63 | 27.21 .03 | 24.52 .17 | 23.99 .02 | 23.53 .08 | 23.95 .14 | 1.93 .02 | — | 1.99 .01 | 27.25 |
| 1985 Jan 27.35 | +57.31 | 0.197 | 3.94 | 27.00 .03 | — | 23.95 .02 | 23.61 .05 | 23.96 .09 | 1.77 .02 | — | 1.87 .01 | 27.04 |
| 1985 Jan 28.43 | +58.39 | 0.198 | 3.79 | — | — | — | — | 24.13 .00 | — | — | 1.94 .01 | — |
| 1985 Feb 15.32 | +76.28 | 0.221 | 3.86 | — | — | 24.04 .03 | — | 23.88 .16 | 1.73 .05 | — | 1.74 .02 | — |
| 2004 Dec 10.17 | –75.89 | 0.208 | 4.40 | 26.79 .13 | 24.81 .13 | 24.18 .06 | 23.92 .07 | 24.31 .05 | 1.38 .24 | 1.23 .12 | 1.20 .11 | 26.82 |
| 2004 Dec 10.20 | –75.87 | 0.208 | 4.59 | 26.97 .08 | 24.72 .13 | 24.25 .05 | 23.87 .09 | 24.36 .04 | 1.22 .32 | 1.10 .16 | 1.06 .14 | 27.00 |
| 2005 Mar 8.15 | +12.09 | 0.138 | 4.48 | 27.07 .06 | 25.18 .05 | 24.59 .02 | 24.04 .07 | 24.54 .04 | 1.73 .13 | 1.74 .06 | 1.64 .07 | 27.13 |
| 2005 Mar 10.17 | +14.11 | 0.139 | 4.48 | 27.12 .05 | 25.03 .06 | 24.60 .02 | 24.03 .07 | 24.54 .04 | 1.90 .09 | 1.56 .06 | 1.71 .06 | 27.18 |
| 2005 Mar 10.20 | +14.14 | 0.139 | 4.27 | 27.21 .10 | 25.17 .08 | 24.64 .08 | 24.21 .10 | 24.58 .05 | 0.64 .54 | 1.80 .06 | 1.76 .06 | 27.27 |
| 2005 Apr 6.17 | +41.11 | 0.160 | 4.32 | 26.85 .08 | 24.90 .12 | 24.37 .04 | 23.84 .13 | 24.27 .09 | 1.24 .36 | 1.44 .12 | 1.72 .07 | 26.90 |
| 2005 Apr 6.19 | +41.13 | 0.160 | 4.53 | 26.90 .08 | 24.85 .09 | 24.34 .04 | 23.98 .11 | 24.22 .08 | 1.34 .29 | 1.69 .06 | 1.60 .08 | 26.95 |

**Notes.**
[a] Production rates, followed by the upper, i.e., the positive uncertainty. The "+" and "-" uncertainties are equal as percentages, but unequal in log-space; the "-" values can be computed.
[b] Continuum filter wavelengths: UV (1984/85) = 3650 Å, UV (2004/05) = 3445 Å; blue (2004/05) = 4450 Å; green (1984/85) = 4845 Å, green (2004/05) = 5260 Å.

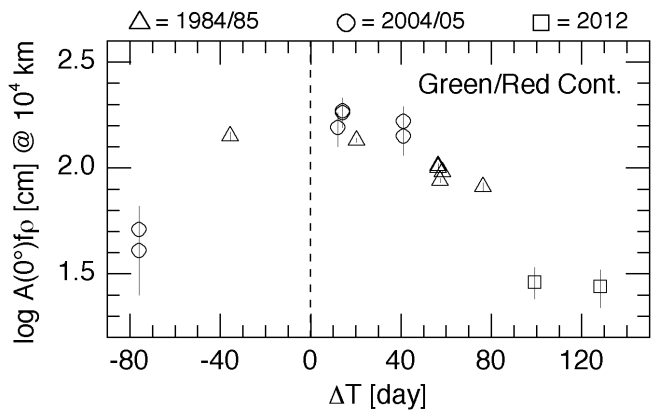

**Figure 11.** Adjusted log $Af\rho$ plotted as a function of time from perihelion. Because of the wide range of solar phase angles encountered throughout the apparitions, 4.6° to 43.2°, $Af\rho$ values have been adjusted to 0° solar phase angle (see values in *Table 4*). In addition, on nights when measurements were made with more than one aperture, $Af\rho$ values always exhibited a decreasing trend with increasing aperture size. Given the nearly order of magnitude range in aperture sizes, we also applied a nominal adjustment to normalize all results to a projected radius of 10,000 km. The result is quite similar in appearance to that of CN shown in *Figure 10*. Finally, we also include measurements extracted from the *R*-band imaging in early 2012 (squares), after first removing the relatively large nucleus contribution. These points confirm the steep drop-off after perihelion.

late 2004/05. Finally, we have attempted to extend the dust measurements even further from perihelion by extracting $Af\rho$ from the *R*-band imaging in 2012. Unlike for 1984/85, where Millis et al. (1988) state that the nucleus was always less than 20% of the measured $Af\rho$ for their 'large' aperture measurements, or 2004/05 when observing circumstance and much larger apertures imply an even smaller amount of nucleus contamination, in 2012 the nucleus was always a major contributor. For the desired aperture of $\rho = 10{,}000$ km, on January 26 the nucleus dominated at about 57% of the total signal, while on February 24 it was about 63%. Therefore, we removed the estimated nucleus contributions before adjusting for solar phase angle, ultimately yielding log $A(0°)f\rho$ values of 1.46 and 1.44, respectively, and these are also plotted in *Figure 11*. (With an even higher contamination in March, coupled with the outburst feature described in *Section 4.1*, $Af\rho$ values would be highly uncertain and are not presented.) These results from 2012 seem exceptionally low, but it is unlikely to be due to the different wavelength for the continuum since dust is generally 'pink' in color, which would yield a higher, not lower value. In any case, it is clear that the strong fall-off in production rates following perihelion continues for at least another six weeks, further supporting our hypothesis that the dominant source region on the nucleus must be moving into winter.

## 5. SUMMARY AND DISCUSSION

We imaged comet 49P/Arend-Rigaux on 33 nights between 2012 January and May and obtained lightcurves of the nucleus. By phasing all of the lightcurves, a synodic rotation period of 13.450 ± 0.005 hr was determined. Similarly, PDM and L-S both yielded a rotation period of 13.452 hr. Rotation periods of monthly and bi-monthly subsets, as determined by inspection, are suggestive of a slight increase in the rotation period during the 2011/12 apparition, consistent with a retrograde rotation of the nucleus. Even though the change of 0.008 hr between January and May is small, and within the calculated uncertainties, it is in agreement with the expected synodic-sidereal offsets.

In order to determine whether the rotation period of 49P/Arend-Rigaux has undergone significant change, we reanalyzed data from the 1984/85 apparition. By combining the observational data from three independent groups, we significantly increased the number of nights of data and thus were able to determine the rotation period to a higher degree of precision. Inspection revealed a period of 13.45 ± 0.01 hr, implying that any change in rotation period was less that 14 s per apparition between 1984/85 and 2011/12. This small change in the rotation period comes as no surprise considering the large size of the nucleus combined with the lack of a detectable jet which could result in a torque. This result further highlights that comet 49P/Arend-Rigaux is largely inactive.

Samarasinha and Mueller (2013) introduced a parameter, $X$, in order to predict changes in rotational periods, that should be approximately constant for comets with similar bulk densities, nucleus shapes, and activity patterns. Their data were limited due to the very low number of comets with both detectable changes in rotation period and reasonable estimates of nucleus size. Using their *Equation 12* and our upper limit for the change of the rotation period of 14 s per orbit and $\zeta_{A-R/Encke}=0.31$ (Samarasinha, personal comm.) we find $X/X_{Encke}<0.45$ where $X$ has been normalized to comet 2P/Encke following the methodology of Samarasinha and Mueller (2013). Although $X/X_{Encke}$ for comet 49P/Arend-Rigaux is lower than the four comets in their sample, the upper limit differs from the most extreme case by only a factor of 4. This further suggests what is intuitively obvious, that such a low change in rotation period may not be unusual given comet 49P/Arend-Rigaux's low activity rate and large nucleus size. Alternatively, the formalism for X is not valid for 49P/Arend-Rigaux as the rotation period might not have changed.

Despite the negligible change in comet 49P/Arend-Rigaux's rotation period over the past three decades, we encourage additional measurements of the rotation period on future apparitions. Comet 49P/Arend-Rigaux was one of the first comets to have its nucleus rotation period determined to high precision, so it offers a nearly unique opportunity to monitor the long-term effects of cometary activity on rotation. The only other comparable object is 10P/Tempel 2, another large, weakly active comet. Comet 10P/Tempel 2's rotation period has been measured on multiple epochs since 1988 (e.g., Jewitt & Meech 1988, Mueller & Ferrin 1996, Knight et al. 2011; 2012), yielding the smallest measured change in rotation period of any comet, ~16 s per orbit (Schleicher et al. 2013). Given that extinct or nearly dormant comets make up a non-negligible fraction of the near-Earth object population (e.g., Mommert et al. 2015), gaining a better understanding of the long term behavior of comets as they become inactive may prove helpful in efforts to assess the risk they pose.

We found an unexpected increase in brightness in 2012 March which was accompanied by a jet-like structure whose appearance evolved over ~2 weeks. By measuring the projected distance of the particles relative to the nucleus, we were able to constrain the grain velocities to a minimum of 17 m s$^{-1}$ and 56 m s$^{-1}$ for the inner and outer ends of the jet-like feature respectively. This allowed us to estimate that the event took place on 2012 March 15 around 18 UT and lasted for no more than 2 hours. Even though this was a short impulse event, we see a jet-like feature presumably due to the particles travelling at a large range of different velocities, resulting in the grains spreading out radially from the nucleus. Whilst we do not believe this event to be a seasonal effect, we encourage observations at the same orbital position.

This outburst is similar to an outburst of 67P/Churyumov-Gerasimenko detected from the ground by Boehnhardt et al. (2016) and confirmed by Knight et al. (2017). Such outbursts are orders of magnitude smaller than the large outbursts of 17P/Holmes (e.g., Montalto et al. 2008, Schleicher 2009) or 29P/Schwassmann-Wachmann 1 (e.g., Roemer 1958, Whipple 1980). Small outbursts are likely common (e.g., A'Hearn et al. 2005), but require frequent, high quality observations to be detected. Appropriate observations are likely to be obtained for large numbers of comets in the near future via the Zwicky Transient Facility (ZTF) and, later, the Large Synoptic Survey Telescope (LSST). We encourage the study of outbursts with ZTF and LSST data as they are likely to yield new insights into the internal composition of comets as well as the processes acting at or near a comet's surface.

The amplitudes of the light curves from the 1984/85 and 2011/12 apparitions were very similar despite the different viewing geometry, implying that we saw the comet at similar sub-Earth latitudes in both apparitions. Furthermore, the large amplitudes of the light curves suggest that we saw the comet near equator-on, at an obliquity of near 0° or near 180°. The apparent small lengthening of the rotation period evident in subsets of the 2012 data implies that the retrograde (180°) solution is correct.

Narrowband photometry of the coma during the 1984/85 and 2004/05 apparitions yielded production rates for a number of species, and showed a strong pre-/post-perihelion asymmetry. Furthermore, the very steep $r_H$-dependence post-perihelion suggests a strong seasonal effect due to a changing sub-solar latitude. This implies that the axis is tilted in an intermediate position, such that the change in amplitude is minimal and yet that the source region is able to change from 'summer' to 'winter' in a short time interval. A similar effect was observed on 9P/Tempel 1, which had both a small tilt as well as strong seasonal effects, made possible by the source region being located very close to the pole (e.g., Schleicher 2007).

The location of the presumed source region on the surface of 49P/Arend-Rigaux is unknown; however, the photometric measurements imply that there are distinct active regions as opposed to uniform leakage across the surface. Furthermore, photometry revealed that comet 49P/Arend-Rigaux is the first comet to show OH and the minor species exhibiting opposite trends. This is also indicative of multiple distinct active regions on the surface. Finally, water production rates, based on OH measurements, showed that comet 49P/Arend-Rigaux has the fifth lowest active fraction in our entire photometric database. This is consistent with the lack of an observable change in nucleus rotation period. Additional gas production rate measurements during future apparitions are highly desirable to investigate the surprising opposite trend of OH and minor species and/or look for evolution of activity as the comet ages.

## ACKNOWLEDGMENTS

We thank Brian Skiff and Larry Wasserman for helping us obtain data and Jessica Sunshine for useful discussions regarding interpretation of our results. Many thanks to Emily Kramer for attempting to obtain light curves from the 2004 data, to Allison Bair for assistance in creating several tables and to Tony Farnham for helping with calculations of the synodic-sidereal effects. We also thank Beatrice Mueller for a thorough and helpful review. Additional thanks go to the University of Sheffield for presenting N.E. with the opportunity of spending an academic year at the University of Maryland. This research made use of Astropy, a community-developed core Python package for Astronomy (Astropy Collaboration, 2013), as well as PyAstronomy. It also used SAOImage DS9, developed by Smithsonian Astrophysical Observatory and the "Aladin sky atlas" developed at CDS, Strasbourg Observatory, France (Bonnarel et al. 2000). M.M.K. and D.G.S. were supported by NASA's Planetary Astronomy grant NNX14AG81G. N.E. was partially supported by the Marcus Comet Fund at Lowell Observatory.

## APPENDIX

As noted in *Section 3.3*, Wisniewski et al. (1986) presented and then published a very short paper in the proceedings of Asteroids, Comets, and Meteors II. In their paper they presented preliminary results from photometric measurements they had obtained of two comets, 28P/Neujmin 1 and 49P/Arend-Rigaux in 1984 and 1985, respectively. Specifically, for comet 49P/Arend-Rigaux they gave the derived period and presented two figures, the first a sample lightcurve of their 4[th] night of observations (1985 January 20 UT), and the second a phased lightcurve using their preferred period for all eight nights of data (January 17, 18, 19, 20, 21, and February 15, 16, 17), with each night having a different symbol. A separate, CCD image of comet 49P/Arend-Rigaux is also presented, showing that the comet had a non-negligible coma, but that the nucleus was readily detected. While an aperture of 12 arcsec was used with the photoelectric photometer to minimize the coma contribution (Wisniewski & Fay 1985), they specifically note that the true amplitude of variability of the nucleus itself is therefore much larger than the measured amplitude due to coma contamination (Wisniewski et al. 1986).

Because no data were tabulated, we had to extract data from the phase plot and compute the UT times associated with each data point based on the period and the zero point used for the phasing, and the knowledge of which night each point was associated with (based on the symbols). Unfortunately, there were several problems with what might have been a straightforward process of deriving the UT times. The first difficulty was that the authors gave different values for the period in the text (1.138 day) than in the phase plot's key (1.134 day). We therefore performed all of our derivations twice, once with each value, until we could determine which value for the period had been used in creating the phase plot (see below). The second problem is that no indication was given as to the date and time to which zero rotational phase corresponded. Finally, as was immediately evident by simply comparing the January 20 lightcurve plot to the same night's data on the phase plot, while the overall shape of the lightcurve was the same, the detailed pattern of points exhibited numerous discrepancies.

After enlarging and scanning both figures, we used a digitization utility to measure each point for a given night, repeating for each of the eight nights on the phase plot; with the magnified view, the identification was ambiguous for only a few overlapping points. For each night we determined the relative number of rotational cycles based on the period, and this was added to the extracted phase value and then multiplied by the period to get a relative time in days. We then compared the derived lightcurve in units of decimal hours for January 20 to the original UT lightcurve for this night, and determined that the data points on the phase plot had non-negligible scatter in both dimensions. While most points were plotted within ±0.01 day of their values on the UT plot, a few differed by more than 0.02 day. However, magnitudes exhibited a systematic shift on average, with the majority shifted lower on the phase plot by 0.01 mag while several others differed by ±0.02 mag or more. Since the points on the original UT plot exhibit a much cleaner pattern and more regular spacing, we conclude that the authors were less careful when plotting the points (presumably by hand) on the phase plot, possibly because this was a preliminary result presented in conference proceedings and not intended for a final, refereed publication.

Because of this 'jitter' introduced in the phase plot, determining the zero point and the period used in its creation was made more difficult but eventually was sorted out using a variety of constraints. Based on the UT plot for January 20, the zero point for phasing was near a value of -0.5 day from UT January 17.0 (observations began on the 17[th]). An offset of exactly -0.50 day would imply Julian Dates had been used, and Faye and Wisniewski (1978) had used 0 hr Julian Date for the zero point in their rotational study of Comet 6P/d'Arrest. This zero point and the shorter period of 1.134 day (listed within the phase plot) both gave matching times (to within 10 minutes, consistent with the jitter) of the lightcurve plot on January 20. Extracted times on all eight nights were also compared to the comet's ephemeris, confirming that for this scenario the comet was always quite accessible; in fact, observations usually started and stopped when the comet reached 50° altitude on either side of the meridian. In contrast, using the longer period (1.138 day) required a zero point offset of about -0.55 day, which does not correspond to a sensible starting point. We also compared the derived lightcurves with those by the Millis et al. (1988) and Jewitt & Meech (1986) for nights in common, and the longer period exhibited an unacceptable systematic drift between the times of lightcurve maxima over the apparition. Having eliminated the 1.138 day solution, we concluded that the last digit in the text was a simple typo, and that the authors had indeed originally phased the data using a period of 1.134 day.

**Table 7**
Dates, times and magnitudes extracted from Wisniewski et al. (1986) for 1985 observations.

| Date | UT | m | Date | UT | m | Date | UT | m | Date | UT | m |
|---|---|---|---|---|---|---|---|---|---|---|---|
| Jan 17 | 6.20 | 6.091 | Jan 19 | 7.75 | 5.983 | Jan 21 | 6.93 | 5.826 | Feb 16 | 5.61 | 6.161 |
| Jan 17 | 6.67 | 6.097 | Jan 19 | 8.09 | 5.981 | Jan 21 | 7.25 | 5.782 | Feb 16 | 6.13 | 6.328 |
| Jan 17 | 6.78 | 6.075 | Jan 19 | 8.47 | 5.979 | Jan 21 | 7.46 | 5.771 | Feb 16 | 6.36 | 6.283 |
| Jan 17 | 7.45 | 6.060 | Jan 19 | 8.54 | 5.941 | Jan 21 | 7.75 | 5.761 | Feb 16 | 6.84 | 6.188 |
| Jan 17 | 7.70 | 6.030 | Jan 19 | 8.84 | 5.951 | Jan 21 | 8.25 | 5.755 | Feb 16 | 7.07 | 6.140 |
| Jan 17 | 8.31 | 5.984 | Jan 19 | 9.21 | 5.941 | Jan 21 | 8.78 | 5.799 | Feb 16 | 7.33 | 5.919 |
| Jan 17 | 8.76 | 5.952 | Jan 19 | 9.29 | 5.922 | Jan 21 | 10.38 | 6.060 | Feb 16 | 7.58 | 5.919 |
| Jan 17 | 8.94 | 5.972 | Jan 19 | 9.47 | 5.948 | Jan 21 | 10.99 | 6.105 | Feb 16 | 7.97 | 5.819 |
| Jan 17 | 9.26 | 5.931 | Jan 19 | 9.67 | 5.923 | Jan 21 | 11.52 | 6.199 | Feb 16 | 8.23 | 5.837 |
| Jan 17 | 9.44 | 5.971 | Jan 19 | 9.89 | 5.945 | Feb 15 | 3.55 | 6.099 | Feb 16 | 8.55 | 5.797 |
| Jan 17 | 9.76 | 5.919 | Jan 19 | 10.10 | 5.960 | Feb 15 | 4.62 | 5.905 | Feb 16 | 8.92 | 5.776 |
| Jan 17 | 10.09 | 5.899 | Jan 19 | 10.49 | 5.988 | Feb 15 | 5.04 | 5.903 | Feb 17 | 2.16 | 6.299 |
| Jan 17 | 10.33 | 5.851 | Jan 19 | 10.80 | 6.018 | Feb 15 | 5.43 | 5.816 | Feb 17 | 2.50 | 6.320 |
| Jan 17 | 10.51 | 5.888 | Jan 19 | 10.73 | 6.064 | Feb 15 | 5.76 | 5.799 | Feb 17 | 2.90 | 6.245 |
| Jan 17 | 10.92 | 5.862 | Jan 19 | 11.08 | 6.095 | Feb 15 | 6.32 | 5.720 | Feb 17 | 2.92 | 6.180 |
| Jan 17 | 11.02 | 5.891 | Jan 19 | 11.35 | 6.159 | Feb 15 | 6.46 | 5.748 | Feb 17 | 3.25 | 6.180 |
| Jan 17 | 11.32 | 5.903 | Jan 20 | 5.85 | 5.897 | Feb 15 | 6.94 | 5.817 | Feb 17 | 3.81 | 6.037 |
| Jan 18 | 9.73 | 6.126 | Jan 20 | 6.32 | 5.946 | Feb 15 | 7.16 | 5.839 | Feb 17 | 4.19 | 6.042 |
| Jan 18 | 9.81 | 6.140 | Jan 20 | 6.53 | 5.925 | Feb 15 | 7.19 | 5.812 | Feb 17 | 4.31 | 6.032 |
| Jan 18 | 10.14 | 6.100 | Jan 20 | 7.12 | 5.931 | Feb 15 | 7.41 | 5.827 | Feb 17 | 5.12 | 5.932 |
| Jan 18 | 10.51 | 6.100 | Jan 20 | 7.07 | 6.024 | Feb 15 | 7.71 | 5.850 | Feb 17 | 5.21 | 5.946 |
| Jan 18 | 10.73 | 6.113 | Jan 20 | 7.22 | 6.087 | Feb 15 | 8.03 | 5.799 | Feb 17 | 5.74 | 5.877 |
| Jan 18 | 10.80 | 6.081 | Jan 20 | 7.30 | 6.075 | Feb 15 | 8.47 | 5.862 | Feb 17 | 5.89 | 5.838 |
| Jan 18 | 10.98 | 6.093 | Jan 20 | 7.85 | 6.208 | Feb 15 | 8.88 | 5.942 | Feb 17 | 6.18 | 5.838 |
| Jan 18 | 11.21 | 6.041 | Jan 20 | 8.28 | 6.226 | Feb 15 | 9.09 | 6.189 | Feb 17 | 6.62 | 5.899 |
| Jan 18 | 11.58 | 6.003 | Jan 20 | 8.33 | 6.265 | Feb 15 | 9.53 | 6.218 | Feb 17 | 7.16 | 5.977 |
| Jan 18 | 11.69 | 5.963 | Jan 20 | 8.95 | 6.235 | Feb 15 | 9.74 | 6.241 | Feb 17 | 7.23 | 5.928 |
| Jan 18 | 11.88 | 5.921 | Jan 20 | 9.32 | 6.189 | Feb 16 | 2.69 | 5.891 | Feb 17 | 7.65 | 6.049 |
| Jan 18 | 12.18 | 5.933 | Jan 20 | 9.96 | 6.112 | Feb 16 | 2.83 | 5.913 | Feb 17 | 7.84 | 6.137 |
| Jan 19 | 5.72 | 6.308 | Jan 20 | 10.44 | 6.070 | Feb 16 | 3.31 | 5.849 | Feb 17 | 8.20 | 6.178 |
| Jan 19 | 5.88 | 6.283 | Jan 20 | 10.40 | 6.027 | Feb 16 | 3.61 | 5.875 | Feb 17 | 8.36 | 6.164 |
| Jan 19 | 6.18 | 6.168 | Jan 20 | 10.47 | 5.986 | Feb 16 | 3.97 | 5.941 | Feb 17 | 9.09 | 6.313 |
| Jan 19 | 6.46 | 6.128 | Jan 20 | 11.81 | 5.962 | Feb 16 | 4.21 | 6.017 | Feb 17 | 9.51 | 6.317 |
| Jan 19 | 6.70 | 6.091 | Jan 20 | 11.97 | 5.945 | Feb 16 | 4.57 | 6.083 | Feb 17 | 10.05 | 6.170 |
| Jan 19 | 7.07 | 6.075 | Jan 21 | 5.96 | 6.055 | Feb 16 | 4.68 | 6.037 | Feb 17 | 10.09 | 6.059 |
| Jan 19 | 7.45 | 6.069 | Jan 21 | 6.41 | 5.886 | Feb 16 | 5.20 | 6.080 | Feb 16 | 5.61 | 6.161 |
| Jan 19 | 7.68 | 6.023 | Jan 21 | 6.73 | 5.839 | Feb 16 | 5.52 | 6.152 | Feb 16 | 6.13 | 6.328 |

The extracted magnitudes and derived decimal dates are listed in *Table 7*. As with January 20, we assume that similar jitter affects all eight nights of data and we assume that similar uncertainties as detailed above are present throughout. However, as with January 20, the ensemble lightcurve on each night should be reasonable, especially when determining the timing of maxima and minima, the critical constraints for period determinations.